\documentclass[aps,pra,twocolumn,showpacs,floatfix]{revtex4-1}
\usepackage{amsmath,amssymb,amsfonts,bbm,graphicx,hyperref}
\def\Tr{\hbox{Tr}}

\newcommand{\bmsigma}{\boldsymbol \sigma} 
\newcommand{\bmLambda}{\boldsymbol \Lambda} \def\D{{\rm D}}
\newcommand{\bmlambda}{\boldsymbol \lambda} 
\newcommand{\bmalpha}{\boldsymbol \alpha} 
\newcommand{\bmY}{\boldsymbol Y} 
\newcommand{\shs}{{{\scriptstyle \!H\!S}}}
\def\X{\boldsymbol{X}}
\def\S{{\sf S}}
\def\Tr{\hbox{Tr}} \def\sigmaCM{\boldsymbol{\sigma}}
\newtheorem{tth}{Theorem}
\newtheorem{lm}{Lemma A$\!\!\!$}
\newtheorem{lmb}{Lemma B$\!\!\!$}
\begin{document}
\title{Quantifying non-Gaussianity for quantum information}
\author{Marco G. Genoni}\email{marco.genoni@fisica.unimi.it}
\affiliation{CNISM, UdR Milano, I-20133 Milano, Italia}
\author{Matteo G. A. Paris}\email{matteo.paris@fisica.unimi.it}
\affiliation{Dipartimento di Fisica, Universit\`a degli Studi di Milano,
I-20133 Milano, Italia}
\date{\today}
\begin{abstract}
We address the quantification of non-Gaussianity of states and
operations in continuous-variable systems and its use in quantum
information. We start by illustrating in details the properties and the
relationships of two recently proposed measures of non-Gaussianity based
on the Hilbert-Schmidt (HS) distance and the quantum relative entropy
(QRE) between the state under examination and a reference Gaussian
state.  We then evaluate the non-Gaussianities of several families of
non-Gaussian quantum states and show that the two measures have the same
basic properties and also share the same qualitative behaviour on most
of the examples taken into account. However, we also show that they
introduce a different relation of order, i.e. they are not strictly
monotone each other.  We exploit the non-Gaussianity measures for states
in order to introduce a measure of non-Gaussianity for quantum
operations, to assess Gaussification and de-Gaussification protocols,
and to investigate in details the role played by non-Gaussianity in
entanglement distillation protocols. Besides, we exploit the QRE-based
non-Gaussianity measure to provide new insight on the extremality of
Gaussian states for some entropic quantities such as conditional
entropy, mutual information and the Holevo bound. We also deal with
parameter estimation and present a theorem connecting the QRE non-Gaussianity 
to the quantum Fisher information. Finally, since evaluation of the QRE 
non-Gaussianity measure requires the knowledge of the full
density matrix, we derive some {\em experimentally friendly} lower
bounds to non-Gaussianity for some class of states and by considering the
possibility to perform on the states only certain efficient or
inefficient measurements.
\end{abstract}
\maketitle
\section{Introduction}\label{s:intro}
In the recent years we have witnessed a big effort in the theoretical
and experimental investigation of continuous-variable (CV) quantum
information. Gaussian states are experimentally produced with an high
degree of control, especially in quantum optics, and Gaussian
measurements may be effectively implemented in different settings.
Besides, despite they belong to an infinite-dimensional Hilbert space,
Gaussian states are easy to handle from the theoretical point of view,
being fully described by the first and second moments of the canonical
operators \cite{Gaussians1,Gaussians2,Gaussians3}. 
The remarkable role of Gaussian states has been highlighted in 
\cite{Wolf}, where it has been 
proved that they are extremal at fixed covariance matrix for several
relevant quantities as channel capacities and entanglement measures 
and 
also in the framework of CV quantum key distribution in \cite{Gro04,Nav06,Raul06}, 
where it has been 
shown that {\em Gaussian attacks} are optimal against all individual and
collective eavesdropping strategies.
For these reasons, Gaussian states played a prominent role in the
development of CV quantum information and, as a matter of fact, most of 
the protocols designed for finite-dimensional Hilbert spaces have been 
firstly translated in the CV setting for Gaussian states \cite{Brau}.  
\par
\begin{figure}[h!]
\includegraphics[width=0.92\columnwidth]{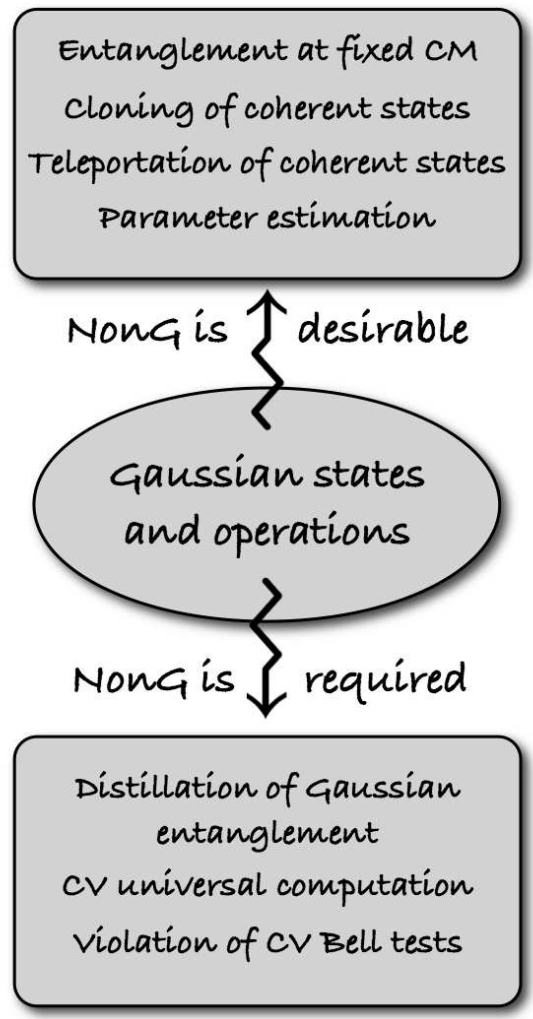}
\caption{The departure from the Gaussian world is required to achieve
specific task such as the distillation of Gaussian entanglement, 
universal quantum computation with Gaussian cluster states and 
the violation of loophole-free Bell tests with continuous variables 
(the lower box) and lead to an increase of entanglement at fixed covariance
matrix, with improvement of relevant protocols
such as teleportation and cloning of coherent states and parameter
estimation of both unitary and lossy channels  (the upper box).
}\label{f:sch}
\end{figure}

In the recent years, however, it has been realized that there are
situations wherein non-Gaussianity (nG) in the form of non-Gaussian
states or non-Gaussian operations is either required or desirable to
achieve some relevant tasks in quantum information processing.  As for
example, it is known that nG is crucial for the realization of
entanglement distillation \cite{nGDist1,nGDist2,nGDist3}, quantum
error correction \cite{nGEC} and cluster states quantum computation
\cite{nGQC1,nGQC2}. Besides, a non-Gaussian measurement and/or
non-Gaussian states are crucial to observe violation of loophole free 
Bell tests with continuous variables
\cite{ban98,ban99,jeo00,fil02,che02,nha,oli05,san05,daf05,Bell32}.  In
addition, improvement of quantum teleportation and quantum cloning of
coherent states can be obtained by using respectively non-Gaussian
states or non-Gaussian operations
\cite{nGTLP1,nGTLP2,nGTLP3,nGClon}. In turn, bipartite Gaussian
states have minimum entanglement for given second moments and this
influences their performances in quantum information protocols.
Non-Gaussian operations also find application in noiseless amplification
\cite{Fer10,Xia10} obtained in conditional fashion, whereas non-Gaussian
states have been proven useful to improve parameter estimation in
quantum optics \cite{EKerr,GG09}. The current state
of the art is schematically depicted in Fig.  \ref{f:sch}.
\par
For the reasons outlined above several protocols have been designed
theoretically \cite{DeGaussTh1,DeGaussTh2} and experimentally realized
\cite{Lvo01,Wen04,Zav04,Our06a,Nee07,Our07a,Zav07a,Par07,Zav09,Our09,Our07} to
produce single mode or two-mode non-Gaussian states, in different
physical settings, and in particular to perform squeezing purification
\cite{PurFiurPRL} and CV entanglement
distillation \cite{Our07a,DistBrowne,EDHageNatPh,DistTaka}.  Basically,
they may be divided into two main categories: those based on nonlinear
interaction of order higher than two \cite{ngOPO1,ngOPO2}, as for
example the Kerr effect \cite{kerr:PRL:01,kerr:PRA:06,kerr:NJP:08}), and
those based on conditional measurements. Indeed, the nonlinear dynamics
induced by conditional measurements has been analyzed for a large
variety of schemes
\cite{eng,dar,oli,pgb,ff,zlc,uni,opq,cla,rob,sab,pla,bno,koz,kim}, also
including, besides photon addition and subtraction schemes, optical
state truncation of coherent states \cite{pgb}, state filtering by
active cavities \cite{ff,zlc}, synthesis of arbitrary unitary operators
\cite{uni} and generation of optical qubit by conditional interferometry
\cite{opq}.  Conditional state generation has been achieved in the low
energy regime~\cite{Our07,DistTaka,ger10} by using single-photon
detectors, and also in the mesoscopic domain \cite{mas06,ps10,MTWBRD}.
Realisations of non-Gaussian states have been also reported in optical
cavities \cite{Del08}, and in superconducting circuits \cite{Hof08}. 
\par
Being recognized as a resource for CV quantum information the need
of quantifying the nG character of states and operations naturally 
arises and different measures of non-Gaussianity have been proposed 
\cite{nGHS, nGRE, nGSimon}.
These measures have been used to  assess the role of nG in different
quantum information and communication tasks as teleportation
\cite{TLPIllu}, quantum estimation \cite{EKerr}, experimental
entanglement quantification \cite{nGEE} and entanglement transfer
between CV states and qubits \cite{nGET1,nGET2}.  In
\cite{nGHud} the relationship between nG and  the Hudson's theorem
\cite{HudsonTh} have been studied, obtaining at fixed purity an upper bound
for non-Gaussian states having a positive Wigner function, while in
\cite{nGUncRel} nG bounded uncertainty relations are derived.  The
entropic measure proposed in \cite{nGRE} has been used to quantify
exactly the nG of experimentally produced photon-added coherent
states and a lower bound  has been evaluated
experimentally in \cite{nGCond} for conditional states obtained via
an inefficient photo-detection on classically correlated thermal beams.  
\par 
In this paper, we address in details the quantification of non-Gaussianity 
of states and operations in continuous-variable systems and analyze its 
use in quantum information. At first, we review the two measures proposed in
\cite{nGHS} and \cite{nGRE} by studying in more details their
properties and their relationships and then exploit them to assess
some relevant Gaussification and de-Gaussification protocols, and to 
investigate the role of non-Gaussianity in entanglement distillation 
and quantum communication.
\par 
The paper is structured as follows: in the next section we briefly
review some topics on the quantification of non-Gaussianity of a
classical probability (density) distribution. In Sec. \ref{s:gauss} 
we introduce notation and give the definition of Gaussian states along 
with their properties,  while in Sec. \ref{s:measures} we review the two quantum
measures of nG, proving their properties and highlighting the
relationships between them. In Sec. \ref{s:zoology} we evaluate
the nG measures for some relevant non-Gaussian states, comparing them
and observing if and when they give the same order relation. In Sec.
\ref{s:GaussDeGauss} we employ the two measures of non-Gaussianity 
to address Gaussification process due to the interaction of 
the system with a bath of harmonic oscillators in the vacuum state 
(i.e. dissipation a zero temperature) and de-Gaussification process 
due either to phase-diffusion or to self-Kerr interaction. 
In Sec. \ref{s:distillation} we study the role of nG in two paradigmatic
examples of entanglement distillation protocols, while in Sec.
\ref{s:entropic} we show how the amount of nG is related to some
entropic-informational quantities as the Holevo bound, conditional
entropy and mutual information. In Sec. \ref{s:qe} we deal with
parameter estimation and present a theorem
relating the non-Gaussianity and the quantum Fisher information. 
Finally, in Sec. \ref{s:bounds} we 
address the experimental evaluation of nG in situations
where state tomography is not available, and present some experimentally
friendly bounds for the estimation of nG of some classes of states. 
Sec. \ref{s:conclusions} closes the paper with some concluding remarks.
%
%
\section{Non-Gaussianity of a classical probability distribution}\label{s:classical}
According to the central limit theorem
the Gaussian distribution is ubiquitous in the description of natural
phenomena. In turn, deviations from the Gaussian behaviour are often the sign
that some interesting phenomenon occurs
\cite{ing1,ing2,ing3,ing4,ing5,ing6}, 
and thus a considerable attention 
has been devoted to the detection and quantification of non-Gaussianity
of a classical distribution. 
Basically, there are two main approaches: The first one is based on the 
evaluation of higher moments of the distribution, in particular the third 
and the fourth central moments, to assess Skeweness and Kurtosis of the 
distribution in comparison
to those of a Gaussian one. The second approach is based 
on the evaluation of the Shannon entropy of the distribution, upon the
fact that Gaussian distributions maxmize it at fixed variance.
More recently, it turned out that non-Gaussianity is relevant in the framework 
of {\em Independent Component Analysis} (ICA) \cite{ICA2000}. 
ICA is a method developed in the last decades in
which the goal is to find a linear representation of non-Gaussian
data so that the components are statistically independent. 
In this method the solution is obtained by the maximisation
of the nG of the components and thus, to accomplish
this goal, different measures of nG have been proposed. 
\par
Let us consider a scalar-valued random variable $Y$ with a 
probability density function $p(y):=P(Y=y)$. 
Its $k$-th {\em central moments} are defined as 
\begin{align}
 E[(Y-\mu)^k] &= \int_{-\infty}^{+\infty} dy\: (y - \mu)^k \: p(y) 
\end{align}
where 
\begin{align}
\mu &= \int_{-\infty}^{+\infty} dy \: y\: p(y) 
\end{align}
is the mean value of the distribution.
We say that $Y$ is Gaussian distributed if its probability density function is
a Gaussian function, that is
\begin{align}
p(y) = \frac{1}{\sqrt{2\pi\sigma^2}} \exp\left\{-\frac{y-\mu}{2\sigma^2}\right\}
\end{align}
where $\sigma^2 = E[(Y-\mu)^2]$ is the second moment of the distribution,
called variance.
In the following we will present the definition of two measures of nG for a
classical scalar-valued random variable.
\subsection{Kurtosis}
The fist ever considered measure of nG has been the Kurtosis, that is 
the fourth-order cumulant defined by the formula
\begin{align}
K(Y) = E[(Y-\mu)^4] - 3 \sigma^2.
\end{align}
$K(Y)$ is zero for a 
Gaussian random variable while for most (but not all) non-Gaussian 
random variables, takes values different from zero. Kurtosis can be positive
or negative. Random variables that have negative Kurtosis are called 
sub-Gaussian and are characterized by a probability density function
with heavy tails. On the other hand, the ones with positive Kurtosis are 
called super-Gaussian and they have tipically a ``flat'' distribution (constant
near the mean value and very small for ``distant'' values of the variable).
Anyway, typically the nG is measured by the absolute value or the 
square of the Kurtosis. While it is relatively simple to evaluate it if the
probability density function is known, Kurtosis
presents some drawbacks when its value has to be estimated from a 
measured sample. Indeed, it can be very sensitive to outliers and its value
may strongly depend on only a few observations in the tails of the distribution.
For these reasons Kurtosis is not considered a robust measure of nG.
\subsection{Negentropy}
Given a discrete random variable $X=\{x_1,\dots, x_n\}$ with a probability distribution
$p(x_i)=P(X=x_i)$, we can define its (Shannon) entropy as
\begin{align}
H(X) = - \sum_i p(x_i) \log p(x_i). 
\end{align}
This definition can be generalized for a continuous-valued random variable $Y$, 
in which case it is called differential entropy:
\begin{align}
{\sf H}(Y) = - \int\! dy\: p(y) \log p(y)\,.
\end{align}
A fundamental result of information theory states that at fixed variance, 
Gaussian variables have the largest entropy. Following this result, one
may define a measure of nG, called negentropy, as
\begin{align}
N(Y) = {\sf H}(G) - {\sf H}(Y)
\end{align}
where $G$ is the Gaussian random variable with the same variance of $Y$.
Due to the above mentioned result, negentropy is always non-negative and
it is equal to zero only for Gaussian random variables. Negentropy is
thus well justified by statistical theory but its computation is
typically very difficult. However, simpler approximations based on
evaluations of moments of the random variable have been introduced and
used for ICA purposes \cite{negAPP}. In the following we will see that
one of the two quantum measures that will be analyzed, even if defined
starting from an another quantity, will result to be the quantum
analogue of the negentropy here presented.
\section{Quantum Gaussian states}\label{s:gauss}
For concreteness, we will use here the quantum optical terminology of
modes carrying photons, but our approach may be equally applied to any bosonic
(CV) system.  Let us consider a system of $n$ modes described by mode
operators $a_k$, $k=1\dots n$, satisfying the commutation relations
$[a_k,a_j^{\dag}]=\delta_{kj}$. A quantum state $\varrho$ of the $n$
modes is fully described by its characteristic function \cite{Glauber2}
$$
\chi[\varrho](\bmlambda) = \Tr[\varrho\,D(\bmlambda)]
$$
where $D(\bmlambda) = \bigotimes_{k=1}^n D_k(\lambda_k)$
is the $n$-mode displacement operator, with $\bmlambda =
(\lambda_1,\dots,\lambda_n)^T$, $\lambda_k \in \mathbbm{C}$, and where
$$
D_k(\lambda_k) =\exp\{\lambda_k a_k^{\dag} - \lambda_k^* a_k \}
$$
is the single-mode displacement operator. 
Analogously, quantum states can be fully described by the Wigner 
function, i.e. the Fourier transform of the characteristic
function:
\begin{align}
W[\varrho](\alpha) = \int \frac{d^{2n}\bmlambda}{\pi^{2n}} 
e^{\bmlambda^*\bmalpha + \bmalpha^*\bmlambda} \chi[\varrho](\bmlambda) 
\end{align}
The canonical operators are given by:
\begin{align}
q_k &= \frac{1}{\sqrt{2}}(a_k + a^{\dag}_k), \nonumber \\
p_k &= \frac{1}{i\sqrt{2}}(a_k - a_k^{\dag}) \nonumber
\end{align}
with commutation relations given by $[q_j,p_k]=i\delta_{jk}$.
Upon introducing the  real vector $\boldsymbol{R}=(q_1,p_1,\dots,q_n,p_n)^T$,
the commutation relations rewrite as
$$
[R_k,R_j] = i \Omega_{kj}
$$
where $\Omega_{kj}$ are the elements of the symplectic matrix
$\boldsymbol{\Omega} = i \bigoplus_{k=1}^n \sigma_2$, $\sigma_2$
being the $y$-Pauli matrix.
The covariance matrix (CM) $\bmsigma\equiv\bmsigma[\varrho]$ and the vector
of mean values $\X \equiv \X[\varrho]$ of a quantum state
$\varrho$ are defined as
\begin{align}
{X}_j &= \langle R_j \rangle \nonumber \\
\sigma_{kj} &= \frac{1}{2}\langle \{ R_k,R_j \}\rangle - \langle R_j
\rangle\langle R_k\rangle  
\end{align}
where $\{ A,B \} = AB + BA$ denotes the anti-commutator, and
$\langle O \rangle = \Tr[\varrho\:O]$ is the expectation value
of the operator $O$.
\par
A quantum state $\varrho_G$ is referred to as a Gaussian state if
its characteristic function or equivalently the Wigner function, 
have a Gaussian form, in the Cartesian notation
\begin{align}
\chi[\varrho_G](\bmLambda) &= \exp \left\{ - \frac{1}{2}
\bmLambda^T \bmsigma \bmLambda + i \X^T \boldsymbol{\Omega} \bmLambda \right\} \\
W[\varrho_G](\bmY) &=  \frac{\exp \left\{ - \frac{1}{2}
(\bmY-\X)^T \bmsigma^{-1} (\bmY-\X)\right\} }
{(2\pi)^n \sqrt{\hbox{Det}[\bmsigma]}}
\end{align}
where $\bmLambda$ and $\bmY$ are  real vectors, 
\begin{align}
\bmLambda &= (\hbox{Re}\,
\lambda_1, \hbox{Im}\,\lambda_1, \dots, \hbox{Re} \,\lambda_n,
\hbox{Im}\,\lambda_n)^T \\
\bmY &= (\hbox{Re}\, \alpha_1, \hbox{Im}\,\alpha_1, \dots, \hbox{Re} \,\alpha_n,
\hbox{Im}\,\alpha_n)^T.
\end{align}
Of course, once the covariance matrix and
the vector of mean values are given, a Gaussian state is fully
determined. As for example, the purity $\mu[\varrho_G]=\hbox{Tr}[\varrho^2_G]$
of a n-mode Gaussian state may be expressed as 
\begin{align}
\mu[\varrho_G]= \frac1{2^n \sqrt{\det \bmsigma}}\:.
\label{pur}
\end{align}
A n-mode Gaussian state can be always 
written as $$\varrho_G = U_S \otimes_{k=1}^n \nu_k (n_k)\,
U_S^{\dag}$$ 
where $\nu_k(n_k)=(1+n_k)^{-1} [n_k/(1+n_k)]^{a_k^\dag
a_k}$ is a single-mode thermal state with 
$n_k=\hbox{Tr}\left[a_k^\dag a_k\, \nu_k (n_k)\right]$ 
average number of photons, and $U_S$ denotes the unitary 
evolution generated by a generic Hamiltonian at most bilinear in the 
mode operators, that is an evolution corresponding to a symplectic 
transformation in the phase-space \cite{williamson}.
Any mapping, either unitary or completely-positive, transforming 
Gaussian states into Gaussian states is a {\em Gaussian operation}.
\par
For a single-mode system the most general Gaussian state can be 
written as
$$
\varrho_G= D(\alpha) S(\zeta)
\nu(n) S^\dag (\zeta) D^\dag(\alpha) ,
$$
$D(\alpha)$ being the
displacement operator and $S(\zeta) = \exp[\frac12 \zeta (a^{\dag})^2
- \frac12 \zeta^* a^2]$ the single-mode squeezing operator with 
$\alpha,\zeta\equiv r e^{i\varphi} \in{\mathbbm C}$. The corresponding covariance matrix
has entries 
\begin{align}
\sigma_{11}&= (n + \frac12)\,
\left[\cosh(2r)-\sinh(2r)\cos(\varphi)\right] \:, \\
\sigma_{22}&= (n + \frac12)\,
\left[\cosh(2r)+\sinh(2r)\cos(\varphi)\right] \:, \\
\sigma_{12}&=\sigma_{21}=(n+\frac12)\,
\sinh(2r)\sin(\varphi) \:.
\end{align}
The Von-Neumann entropy $\S(\varrho) = - \Tr[ \varrho \: \log
\varrho]$ of a single-mode Gaussian states may be written 
as 
\begin{align}
\S(\varrho_G) & = h(\sqrt{\det \bmsigma}) = h (\frac1{2 \mu}) = h
(n+\frac12) \notag \\ &= (n+1) \log (n+1) - n \log (n) 
\label{VNE}\:,
\end{align}
where we have introduced the function 
\begin{align}\label{funh}
h(x) = (x+\frac12) \log (x+\frac12)- (x-\frac12) \log (x-\frac12)\:.
\end{align}
\par
For a two-mode Gaussian state, the covariance matrix is a real 
is a real $4\times4$ symmetric definite positive block matrix 
with ten independent parameters
\begin{equation}
\boldsymbol{\sigma }=\left( 
\begin{array}{c|c}
A & C \\ \hline
C^{T} & B%
\end{array}%
\right)  \label{CM}
\end{equation}%
Matrices $A$, $B$
and $C$ are $2\times 2$ real matrices, representing respectively the
autocorrelation matrices of the two modes and their mutual correlation
matrix. Any two-mode CM ${\boldsymbol{\sigma }}$ may be brought to 
its {\em standard form} local symplectic operations, i.e. local 
Gaussian operations. In the standard from, matrices $A$ and $B$ are
proportional to the identity and $C$ is diagonal. Using the four local
symplectic invariants $I_{1}\equiv \det (A)$, $I_{2}\equiv \det
(B)$, $I_{3}\equiv \det (C)$, $I_{4}\equiv \det ({%
\boldsymbol{\sigma }})$ the symplectic eigenvalues, denoted by $d_{\pm }$ with 
$d_{-}\le d_{+}$ read as follows
\begin{equation}
d_{\pm }=\sqrt{\frac{\Delta ({\boldsymbol{\sigma }})\pm \sqrt{\Delta ({%
\boldsymbol{\sigma }})^{2}-4I_{4}}}{2},}  \label{eigenvalues sympletic}
\end{equation}%
where $\Delta (\boldsymbol{\sigma })\equiv I_{1}+I_{2}+2I_{3}$.
Using the symplectic eigenvalues, the uncertainty relation re-writes 
as $d_{-}\geq 1/2$ and the Von-Neumann entropy as \cite{ser04}
\begin{align}
\S(\varrho_G) = h(d_-) + h(d_+)\:. 
\end{align}
\section{Quantum nG measures: definitions and properties} \label{s:measures}
In this section we review the definitions of the nG measures for
quantum states proposed in \cite{nGHS, nGRE} and illustrate in
details their properties. Although the two measure are based on
different quantities, as the Hilbert-Schmidt distance and the 
quantum relative entropy, they share the same basic idea: one wants to 
quantify the non-Gaussianity of a quantum state $\varrho$ 
in terms of the distinguishability of the state itself from a reference
Gaussian state $\tau$, chosen as the Gaussian state with the same first
and second moments of $\varrho$, that is such that
\begin{align}
X[\tau] &= X[\varrho] \nonumber \\
{\bf \sigma}[\tau] &={\bf \sigma}[\varrho]\,.\label{eq:RefGauss}
\end{align}
Notice that a similar line of reasoning has been adopted in 
Refs. \cite{Dodonov1,Dodonov2,Mar04} to define a measure of non-classicality 
via the Hilbert-Schmidt distance.
Here, roughly speaking, the two nG measures provide the quantization of the
classical approaches to asses non-Gaussianity based on moments and
negentropy respectively.  In the following we review their properties
and also provide a critical comparison with another quantities proposed
in literature \cite{nGSimon}.
\subsection{Measuring the non-Gaussianity using Hilbert-Schmidt distance
from a Gaussian reference}
Given two quantum states $\varrho_1$ and $\varrho_2$,
the Hilbert-Schmidt distance is defined as
\begin{align}
\D_{\shs}[\varrho_1,\varrho_2] &= \left( \frac{1}{2} 
\Tr[(\varrho_1 - \varrho_2)^2] \right)^{1/2} \nonumber \\
&= \left( \frac{ \mu[\varrho_1] + \mu[\varrho_2] 
-2\kappa[\varrho_1,\varrho_2]}{2} \right)^{1/2} \nonumber 
\end{align}
where $\mu[\varrho]$ is the purity of $\varrho$ and 
$\kappa[\varrho_1,\varrho_2]=\Tr[\varrho_1\:\varrho_2]$
denotes the overlap 
between $\varrho_1$ and $\varrho_2$. 
We define the degree of non-Gaussianity of the state $\varrho$ as
the squared renormalized HS distance \cite{nGHS} 
\begin{align}
\delta_A [\varrho] = \frac{ \D_{\shs}^2[\varrho,\tau]}{\mu[\varrho]} 
\label{eq:nG1}
\end{align} 
of the state $\varrho$ from the state $\tau$, which is a reference 
Gaussian state chosen as in Eq.~(\ref{eq:RefGauss}).
The relevant properties of $\delta_A[\varrho ]$
are summarized by the following Lemmas:
\begin{lm}
$\delta_A[\varrho]=0$ iff $\varrho$ is a Gaussian state.
\end{lm}
{\bf Proof}: If $\delta_A[\varrho]=0$ then $\varrho=\tau$ and thus it is
a Gaussian state. If $\varrho$ is a Gaussian state, then it is uniquely
identified by its first and second moments and thus the reference Gaussian
state $\tau$ is given by $\tau=\varrho$, which, in turn, leads to
$\D_{\shs}[\varrho,\tau]=0$ and thus to $\delta_A[\varrho]=0$. $\square$
\begin{lm}
If $U$ is a unitary map corresponding to a symplectic
transformation in the phase space, i.e. if $U=\exp\{-i H\}$ with
 hermitian $H$ and at most bilinear in the field operators, then
$\delta_A[U\varrho\:U^{\dag}] = \delta_A[\varrho]$.
\end{lm}
{\bf Proof}:  Let us consider $\varrho^\prime = U\varrho\: U^\dag$.
Then the
covariance matrix transforms as $\sigmaCM[\varrho^\prime] =
\Sigma\: \sigmaCM[\varrho]\: \Sigma^T$, $\Sigma$ being the symplectic
transformation associated to $U$. At the same time the vector of
mean values simply translates to $\X^{\prime}=\X+\X_0$, where $\X_0$
is the displacement generated by $U$. Since any
Gaussian state is fully characterized by its first and second
moments, then the reference state must necessarily transform as
$\tau^\prime = U \tau\: U^\dag$, \emph{i.e.} with the same unitary
transformation $U$. Since the Hilbert-Schmidt distance and the purity
of a quantum state are invariant under unitary transformations
the lemma is proved. $\square$
\par
This property ensures that single-mode displacement and squeezing operations, 
as well as two-mode evolutions as those induced by a beam splitter or a
parametric amplifier, do not change the Gaussian character of a quantum 
state. The lemma also allows us to always consider state with zero mean
values.
\begin{lm}
$\delta_A[\varrho]$ is proportional to the squared
$L^2(\mathbbm{C}^n)$ distance between the characteristic functions 
(or alternatively the Wigner functions) of
$\varrho$ and of the reference Gaussian state $\tau$.
In formula:
\begin{align}
\delta_A[\varrho] &\propto \int d^{2n}{\bmlambda} \: \left[
\chi[\varrho](\bmlambda) - \chi[\tau](\bmlambda)\right]^2 \:, \\
\delta_A[\varrho] &\propto \int d^{2n}{\bmalpha} \: \left[
W[\varrho](\bmalpha) - W[\tau](\bmalpha)\right]^2 \:.
\end{align}
\end{lm}
{\bf Proof}: Using the identities
\begin{align}
\Tr[O_1O_2] &= \int \frac{d^{2n}{\bmlambda}}{\pi^n}
\chi[O_1](\bmlambda)\,\chi[O_2](-\bmlambda)\, \\
&= \pi^n \int d^{2n}{\bmalpha}
W[O_1](\bmalpha)\,W[O_2](\bmalpha)\,
\end{align}
and the fact the characteristic functions of self-adjoint operators are
even functions of $\lambda$ we obtain
\begin{align}
{\D}_{\shs}^2[\varrho,\tau] &= \frac{1}{2} \int \frac{d^{2n}{\bmlambda}}{\pi^n}\,
\left[ \chi[\varrho](\bmlambda) - \chi[\tau](\bmlambda) \right]^2\: \\
&= \frac{\pi^{n}}{2} \int d^{2n}{\bmlambda}\, 
\left( W[\varrho](\bmalpha) - W[\tau](\bmalpha) \right)^2\:, 
\end{align}
which proves the Lemma. $\square$.
\par
Since the notion of Gaussianity of a quantum state is connected
to the shape of its characteristic (Wigner) function, and since the
characteristic function of a quantum state belongs to the
$L^2(\mathbbm{C}^n)$ space \cite{Glauber2}, we address
$L^2(\mathbbm{C})$ distance to as a good indicator for the non
Gaussian character of $\varrho$.
\begin{lm}
Consider a bipartite state $\varrho=\varrho_A\otimes\varrho_G$.
If $\varrho_G$ is a Gaussian state then $\delta_A[\varrho] = \delta_A[\varrho_A]$.
\end{lm}
{\bf Proof}: we have
\begin{align}
\mu[\varrho]&= \mu[\varrho_A] \mu[\varrho_G] \nonumber \\
\mu[\tau]&= \mu[\tau_A] \mu[\tau_G] \nonumber \\
\kappa[\varrho,\tau]&=\kappa[\varrho_A,\tau_A]\kappa[\varrho_G,\varrho_G]
\:. \nonumber
\end{align}
Therefore, since $\kappa[\varrho_G,\varrho_G]=\mu[\varrho_G]$ we arrive at
\begin{align}
\delta_A[\varrho] &=
\frac{\mu[\varrho_A] \mu[\varrho_G] + \mu[\tau_A]\mu[\varrho_G]
- 2 \kappa[\varrho_A,\tau_A]\kappa[\varrho_G,\varrho_G]}{2 \mu[\varrho_A]
\mu[\varrho_G]} \nonumber \\
&= \delta_A[\varrho_A] \:.\: \square
\end{align}
\par\noindent
Notice, however, that $\delta_A [\varrho]$ is not generally additive (nor
multiplicative) with respect to the tensor product. If  we
consider a (separable) multi-partite quantum state  in the product form
$\varrho = \otimes_{k=1}^n \varrho_k$,  the non-Gaussianity is
given by
\begin{equation}
\delta_A[\varrho] = \frac{ \prod_{k=1}^n \mu[\varrho_k] +
\prod_{k=1}^{n}\mu[\tau_k]  - 2 \prod_{k=1}^n \kappa[\varrho_k,\tau_k]} {2
\prod_{k=1}^n \mu[\varrho_k] } \label{eq:nGMultiPartite}
\end{equation}
where $\tau_k$ is the Gaussian state with the same moments of
$\varrho_k$.  In fact, since the state $\varrho$ is factorisable,
we have that the corresponding Gaussian $\tau$  is a factorisable
state too. 
\par
For single-mode quantum states we have collected several numerical 
evidences that $\delta_A[\varrho]=1/2$ represents an upper bound for the
HS nG of any quantum state \cite{TRS}. The same conclusion is
indirectly suggested by the results obtained in \cite{nGHud} and
this leads to formulate the following
\par\noindent $ $ 
\par\noindent
{\bf Conjecture A5} {\em For single-mode quantum states we
have that $\delta_A[\varrho]\leq \frac12$.}
\par\noindent $ $ 
\par\noindent
In particular, the conjecture has been numerically verified for 
single-mode CV states expressed as finite superposition 
of Fock number states, i.e. for truncated states of the form 
$\varrho=\sum_{n,k=0}^N \varrho_{nk} |n\rangle\langle k|$.
We have generated at random a large number states for varius values of the
truncating dimension $N$ and evaluated the corresponding nG 
$\delta_A$. Results have shown that the value of the nG $\delta_A$ is bounded 
by $1/2$ and the typical nG (the value of $\delta_A$ with the largest 
occurence) decreases with both the purity and the
truncating dimension.
\subsection{Measuring the non-Gaussianity using the quantum relative 
entropy to a reference Gaussian}
Given two quantum states $\varrho_1$ and $\varrho_2$, 
the quantum relative entropy (QRE) is defined as
\begin{align}
\S(\varrho_1 \rVert \varrho_2) = \Tr[\varrho_1 (\log\varrho_1 - \log\varrho_2)]
\end{align}
As for its classical counterpart, the Kullback-Leiber divergence, 
it can be demonstrated that $0\leq \S(\varrho_1 \rVert \varrho_2) < \infty$
when it is definite, i.e. when the support of the first state in the Hilbert
space $\mathrm{supp} \: \varrho_1 \subseteq \: \mathrm{supp} \: \varrho_2$, is
contained in that of the second one.
In particular, $\S(\varrho_1 \rVert \varrho_2)=0$ iff $\varrho_1\equiv\varrho_2$.
This quantity, though not defining  a proper metric in the Hilbert
space (it is not simmetric in its arguments), 
has been widely used in different fields of quantum information
as a measure of statistical distinguishability for quantum states
\cite{VedralRelEnt, SchumacherRelEnt} because of its nice
properties and statistical meaning. In fact if we consider two 
quantum states $\varrho$ and $\tau$ and we suppose
to perform $N$ measurements on $\varrho$, the probability
of confusing $\varrho$ with $\tau$ is (for large $N$) 
$P_N(\varrho \rightarrow \tau) \sim \exp\{-N \S(\varrho\rVert \tau)\}$. 
\par
The degree of non-Gaussianity of
a state $\varrho$ may be quantified as \cite{nGRE}
\begin{align}
\delta_B[\varrho] &= \S(\varrho \rVert \tau) \nonumber 
\end{align}
where $\tau$ is the reference Gaussian state with the same
first and second moments as in Eq. (\ref{eq:RefGauss}).
Notice that, because of the choice of $\tau$,
$\Tr[\tau \log \tau]= \Tr[ \varrho \log \tau]$ and thus 
\begin{align}
\delta_B[\varrho] &= \Tr[\varrho \log \varrho] - \Tr[\varrho \log \tau] 
= \S(\tau) - \S(\varrho)\,,
\end{align}
where $\S(\varrho)$
is the von Neumann entropy of a quantum state $\varrho$.
The nG measure $\delta_B$ may be considered as the
quantum analogue of the negentropy introduced in Sec. \ref{s:classical}, 
where the differential entropy is replaced by the von Neumann
entropy of the quantum states under investigation.
\par
At fixed Von-Neumann entropy nG is determined by the first two moments
of the canonical operators, which in turn uniquely determine the
reference Gaussian state. Upon using formulas from Sect. \ref{s:gauss} we may
write explicit formulas of $\delta_B$ for single- and two-mode 
states
\begin{align}
\delta_B[\varrho] & = h (\sqrt{\det \bmsigma}) - \S(\varrho) \quad
\hbox{single-mode states}\,, \\
\delta_B[\varrho] & = h (d_-) + h(d_+) - \S(\varrho) \quad
\hbox{two-mode states}\,,
\end{align}
where $d_\pm$ are the symplectic eigenvalues of the two-mode 
CM and $h(x)$ is given  in Eq. (\ref{funh}).
\par
The relevant properties of $\delta_B[\varrho]$ are 
summarized by the following lemmas.
As a matter of fact, the QRE measure of non-Gaussianity owns 
all the relevant properties proved for $\delta_A$, and shows 
additional properties concerning the evolution under generic 
(not unitary) Gaussian maps and under tensor product. 
\begin{lmb}
$\delta_B[\varrho]=0$ iff $\varrho$ is a Gaussian state.
\end{lmb}
{\bf Proof}: If $\delta_B[\varrho]=0$ then $\varrho=\tau$ and thus it is
a Gaussian state. If $\varrho$ is a Gaussian state, then it is uniquely
identified by its first and second moments and thus the reference Gaussian
state $\tau$ is given by $\tau=\varrho$, which, in turn, leads to
$\S(\varrho\rVert\tau)=0$ and thus to $\delta_B[\varrho]=0$. $\square$
\begin{lmb}
If $U$ is a unitary map corresponding to a symplectic
transformation in the phase space, \emph{i.e.} if $U=\exp\{-i H\}$ with
 hermitian $H$ that is at most bilinear in the field operators, then
$\delta_B[U\varrho\:U^{\dag}] = \delta_B[\varrho]$.
\end{lmb}
This property ensures that single-mode displacement and squeezing operations, 
as well as two-mode evolutions as those induced by a beam splitter or a
parametric amplifier, do not change the Gaussian character of a quantum 
state. The lemma also allows us to always consider state with zero mean
values.
\\ \par\noindent
{\bf Proof}: The lemma follows from the invariance of QRE
under unitary operation. $\square$
\begin{lmb}
$\delta_B$ is additive for factorized states:
$\delta_B[\varrho_1\otimes\varrho_2]=\delta_B[\varrho_1] + 
\delta_B[\varrho_2]$. As a corollary we have that if
$\varrho_2$ is a Gaussian state, then $\delta_B[\varrho]=
\delta_B[\varrho_1]$.
\end{lmb}
{\bf Proof}: The overall reference Gaussian state is 
the tensor product of the relative reference Gaussian
states of $\varrho_1$ and $\varrho_2$, $\tau=\tau_1
\otimes\tau_2$. The lemma follows from the additivity of
QRE and the corollary from Lemma B1. $\square$
\begin{lmb}
$\delta_B$ monotonically decreases under partial trace, that is, 
given a bipartite state $\varrho$, then  
$\delta_B[\varrho_A] \leq \delta_B[\varrho]$ and
$\delta_B[\varrho_B] \leq \delta_B[\varrho]$, where
$\varrho_A = \hbox{Tr}_B [\varrho]$,
$\varrho_B = \hbox{Tr}_A [\varrho]$.
\end{lmb}
{\bf Proof} Let us consider the partial trace state $\varrho_A$
($\varrho_B$): its 
CM is the submatrix of $\sigmaCM[\varrho]$ obtained by dropping 
lines and rows involving expectation values on the system $B$ ($A$).
Analogously, the first moment vector is the proper subvector of $\X[\varrho]$. 
Therefore, the reference Gaussian state $\tau_A$ ($\tau_B$) 
must necessarily satisfies $\tau_A =\Tr_B[\tau]$, where $\tau$ is
the Gaussian reference of $\varrho$ ($\tau_B =\Tr_A[\tau]$). 
The QRE monotonically decreases under partial trace and thus 
the lemma is proved.
$\square$ \par
Actually, the above statement can be strenghtened, as it is expressed
by the following Lemma.
\begin{lmb}
Let us given a generic bipartite state $\varrho$,
we have $\delta[\varrho] \geq \delta[\varrho_{A}] + \delta[\varrho_{B}]$. 
\end{lmb}
{\bf Proof}: It has been shown in \cite{QRECoarseG} that QRE decreases monotonically
under a generic (non-linear) coarse graining. A simple example of non-linear
coarse graining that can not be obtained via a completely positive
quantum map is the operation $\varrho \rightarrow \varrho_{A}\otimes
\varrho_{B}$. Because of this property we have $\delta[\varrho]\geq 
\delta[\varrho_A \otimes \varrho_B] = \delta[\varrho_A] + \delta[\varrho_B]$
where in the last equality we have used lemma B3. $\square$
\begin{figure}[h!]
\includegraphics[width=0.48\textwidth]{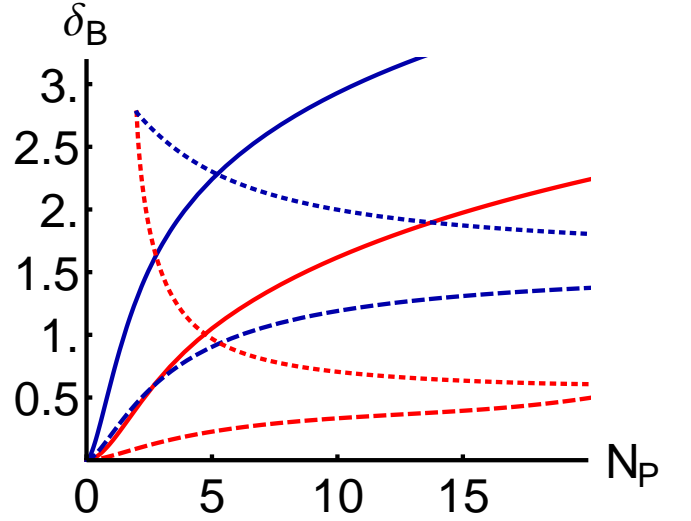}
\caption{
\label{f:1TMCvsPT}
Non-Gaussianity $\delta_B[\psi_{\scriptstyle P}]$ of PNES states (blue upper curves) 
and of their partial traces $2\delta_B[\varrho_{\scriptstyle P}]$ (red lower curves) 
as a function of the total energy of the PNES. Solid curves are for TMC, dashed 
for PSSV and dotted for PASV.}
\end{figure} 
\par
As an example let us consider the class of 
pure two-mode photon-number entangled states (PNES)
\begin{align}
|\psi_{{\scriptstyle P}}\rangle\rangle= 
\sum_n \psi_n\:|n\rangle |n\rangle
\end{align}
together with their (equal) partial traces $\varrho_A$ and $\varrho_B$, 
that is the diagonal mixtures of 
Fock states given by 
\begin{align}
\varrho_{\scriptstyle P} = \sum_n |\psi_n|^2\:|n\rangle\langle n| 
\label{eq:Poiss}.
\end{align} 
Relevant examples of non-Gaussian PNES are given by 
the photon subtracted
(PSSV) $\psi_n  \propto (n+1) x^{n+1}$ and the photon-added
two-mode squeezed vacua (PASV)
$\psi_n  \propto n x^{n-1}$, which are obtained from the 
Gaussian PNES
$\psi_n \propto x^n$
$0 \leq x < 1$ (twin-beam state)
by the
experimentally feasible operations of photon
subtraction $\varrho \rightarrow a_1 a_2\varrho a_1^\dag a_2^\dag$ and
addition $\varrho\rightarrow a_1^\dag a_2^\dag\varrho a_1 a_2 $
respectively \cite{Wen04};
(iii) the pair-coherent or two-mode coherently correlated
states  (TMC) ~\cite{AGA,USE1,USE2} with Poissonian profile
$\psi_n \propto \frac{\lambda^n}{n!}$, $\lambda\in{\mathbb R}$.
The mean energy of PNES  is $N_{\scriptstyle P}=\langle\langle\psi| a_1^\dag a_1 + a_2^
\dag a_2 |\psi\rangle\rangle \equiv 2 N$ where $N=\sum_{n=0}^{\infty}
\lvert \psi_n \lvert^2 n$, whereas correlations between the modes can be
quantified by $C=\mbox{Re}\sum_{n=0}^{\infty} \psi_n^{\ast}
\psi_{n+1} (n+1)$ and entanglement is given by the Von-Neumann entropy
of the partial traces $\epsilon_0=-\sum_n \psi_n^2 \log \psi_n^2$.
In turn, the covariance matrix of a PNES equals that
of a symmetric Gaussian state in standard form, with diagonal elements
equal to $N+\frac12$ and off-diagonal blocks given by
$\boldsymbol{C}=\hbox{diag}(C,-C)$. 
In Fig. \ref{f:1TMCvsPT} we report the nG $\delta_B[\psi_{\scriptstyle
P}]$ of PNESs as a function of the overall energy $N_{\scriptstyle P}$
together with the sum of the nGs of the partial traces i.e.
$\delta_B[\varrho_A]+\delta_B[\varrho_b]=2\delta_B[\varrho_{\scriptstyle
P}]$.  As predicted by the previous lemmas, the nG of a PNES state is always 
larger than the one of its partial traces and also of their sum.
\begin{lmb}
$\delta_B[\varrho]$ monotonically decreases
under Gaussian quantum channels, that is $\delta[\mathcal{E}_G
(\varrho)] \leq \delta[\varrho]$. 
\end{lmb}
{\bf Proof}: Any Gaussian quantum channel can be written as
$\mathcal{E}_G(\varrho) =
\Tr_E [ U_{b} (\varrho\otimes\tau_E) U^{\dag}_{b}]$,
where $U_{b}$ is a unitary operation corresponding to an Hamiltonian
at most bilinear in the field modes and where $\tau_E$ is a Gaussian state
\cite{GaussChannel}. Then, by using lemmas B2, B3 and
B4 we obtain
$\delta_B[\mathcal{E}_G(\varrho)] \leq
\delta[U_{b} (\varrho\otimes\tau_E) U^{\dag}_{b}] = \delta_B[\varrho]$.
$\square$.\par\noindent
In turn, this lemma provides a necessary condition for a
channel to be Gaussian: given a quantum channel $\mathcal{E}$, and a
generic quantum state $\varrho$, if the inequality
$\delta_B[\mathcal{E}(\varrho)]\leq \delta_B[\varrho]$ is not fulfilled, the
channel is nG.
It is also worth to notice that the monotonicity is fulfilled only for a 
proper completely positive (CP) map.
As we will see in Sec. \ref{s:distillation}, even if  we consider the conditional evolution 
corresponding to a Gaussian measurement operator, the nG of the output 
states may increase. Indeed, in this case, we do not consider a full CP-map, but only
one Krauss operator corresponding to the chosen (Gaussian) measurement operator.
\begin{lmb}
For a set of states $\{\varrho_k\}$
having the same first and second moments, then
nG is a convex functional,
that is 
$$
\delta_B[\sum_k p_k \varrho_k]\leq \sum_k p_k \delta_B[\varrho_i],
$$
with $\sum_k p_k =1$. 
\end{lmb}
{\bf Proof}: The states $\varrho_k$, having the same
first and second moments, have the same reference Gaussian
state $\tau$ which in turn is the reference Gaussian state
of the convex combination $\varrho=\sum_k p_k \varrho_k$.
Since conditional entropy $\S(\varrho\rVert\tau)$
is a jointly convex functional respect to both states, we have
$\delta_B[\sum_k p_k \varrho_k]=\S(\sum_k p_k \varrho_k \rVert \tau)
\leq \sum_k p_k \S(\varrho_k \rVert \tau) = \sum_k p_k \delta[\varrho_k]$.
$\square$ \par\noindent
Notice that, in general, $\delta_B$ is not convex, as it may easily
proved upon considering the convex combination of two Gaussian states
with different parameters. 
\begin{lmb}
At fixed average number of photons 
$N = \langle a^{\dag} a \rangle$, the maximum value of 
nG measured by $\delta_B$ for single mode states 
is achieved by pure superpositions of Fock states 
$|\psi_{N}\rangle = \sum_k \alpha_k |n+l_k\rangle$
where $n\geq 0$, $l_k \geq l_{k-1}+3$ or $l_k=0$, and
with the constraint $N= (\det\bmsigma[\nu(N)])^{\frac12}
-\frac12 = n + \sum_k |\alpha_k|^2l_k$.
\end{lmb}
{\bf Proof}: Since $\delta_B[\varrho] = \S(\tau) - 
\S(\varrho)$ we have to
maximize $\S(\tau)$ and, at the same time, minimize $\S(\varrho)$.
For a single-mode system the most general
Gaussian state can be written as $\varrho_G= D(\alpha) S(\zeta)
\nu(n_t) S^\dag (\zeta) D^\dag(\alpha)$, $D(\alpha)$ being the
displacement operator, $S(\zeta)$ the squeezing operator, $\alpha,\zeta \in
{\mathbbm C}$, and $\nu(n_t)$ 
a thermal state with $n_t$ average number of photons.
Displacement and squeezing applied to thermal states increase the
overall energy, while entropy is an increasing monotonous function of
the number of thermal photons $n_t$ and is invariant under
unitary operations. Thus, at fixed energy, $\S(\tau)$ is maximized for
$\tau=\nu(N)$. Therefore, the state with the maximum amount of
nG must be a pure state (in order to have $\S(\varrho)=0$)
with the same CM $\bmsigma=(N+\frac12) {\mathbbm I}$
of the thermal state $\nu(N)$. One can easily check now
that the state with this property is the one indicated 
in the Lemma. One can also observe that by choosing 
$n=N$ and $l_k=0$, we obtain that Fock states $|N\rangle$
are maximum nG states at fixed energy. $\square$
\\
\par\noindent
As it will be clear from the examples presented in the next sections 
the two nG measures measures induce different ordering on the set 
of quantum states, that is we may find a pair of states $\varrho_1$
and $\varrho_2$ such that $\delta_A[\varrho_1]>\delta_A[\varrho_2]$
and $\delta_B[\varrho_1]<\delta_B[\varrho_2]$, or viceversa. 
One may conjecture that, as it happens for entanglement measures 
\cite{EntShash}, we indeed do not have a unique nG 
measure and that different measures correspond to different
operational meanings. As also remarked in Sec. \ref{s:entropic}, an 
operational meaning for $\delta_B$ may be found in terms of
information-theoretic quantities, while an operational meaning for 
$\delta_A$, besides its connection with the distance in the phase-space, 
is still missing.
The two measures are connected each other by means of the inequality
$S(\varrho \rVert \tau) \geq \D_{\shs}^2[\varrho,\tau]$ \cite{OhyaPetz},
which, in turn, implies the inequality 
\begin{align}\label{xxx}
\delta_B[\varrho] \geq \delta_A[\varrho]\: \mu[\varrho]\:.
\end{align}
For pure states (\ref{xxx}) reduces to
$\delta_B[\varrho] \geq \delta_A[\varrho]$.
\subsection{A measure of nG based on the Wehrl entropy}
\begin{figure}[h!]
\includegraphics[width=0.9\columnwidth]{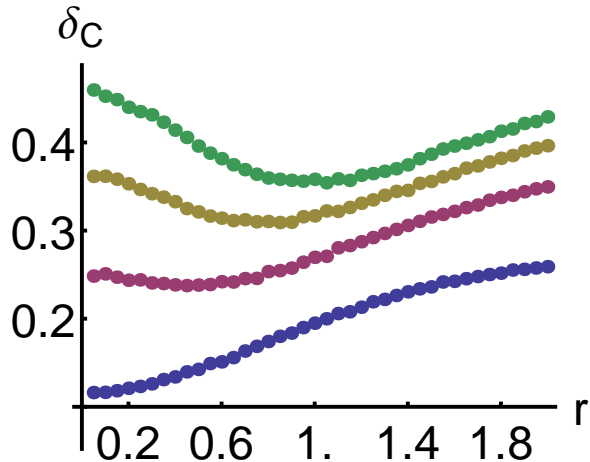}
\caption{
\label{f:2Simon}
The Wehrl entropy-based non-Gaussianity $\delta_C[S(r)|n\rangle
\langle n| S(r)^\dag]$ for squeezed 
Fock states  as a function of the squeezing parameter
$r$. From bottom to top the nG for $n=1,2,3,4$. }
\end{figure}

A different measure of nG has been proposed in \cite{nGSimon}, 
based on the difference between the Wehrl entropies
of the reference Gaussian state and the quantum state in exam:
\begin{align}
\delta_C[\varrho] = H_W(\tau) - H_W(\varrho)
\end{align} 
where 
\begin{align}
H_W(\varrho) = -\! \int_{\mathbbm C}\! d^2 \alpha \:
Q_\varrho (\alpha) \log \left[ \pi\, Q_\varrho(\alpha) \right] 
\end{align}
is the Wehrl entropy, i.e. the differential entropy of the normalized
$Q$-Husimi function $$Q_\varrho(\alpha)=\frac1\pi \langle 0 |D(\alpha)^\dag\varrho 
\:D(\alpha)|0\rangle$$ of the state $\varrho$.
The quantity $\delta_C$ owns reasonable properties in the phase-space,
which are inherited from those of the the $Q$-function. However, it lacks 
an operational meaning and turns to be not invariant under Gaussian 
unitary operations. In order to illustrate this behaviour we have 
(numerically) evaluated 
the non-Gaussianity $\delta_C$ for Fock number states $|n\rangle$ 
subjected to squeezing. In Fig. \ref{f:2Simon} we show 
$\delta_C[S(r)|n\rangle\langle n|S(r)^{\dag}]$ as a function of $r$
for different values of $n$.  
As it is apparent from the plot, the nG is neither
constant nor monotone with the squeezing parameter $r$. 
\subsection{Non-Gaussianity of a quantum operation}
Once one has at disposal a good measure for the non-Gaussianity of 
quantum state this may be exploited to define a measure for the 
non-Gaussian character of a quantum operation. Let us denote by 
${\cal G}$ the whole set of Gaussian states. A convenient definition 
for the non-Gaussianity of a map ${\cal E}$ reads as follows
$$\delta[{\cal E}] = \max_{\varrho \in {\cal G}} \delta [{\cal E}
(\varrho)]\,,$$ where ${\cal E} (\varrho)$ denotes the quantum state
obtained after the evolution imposed by the map. Indeed, we have 
$\delta[{\cal E}_g]=0$ iff ${\cal E}$ is a Gaussian map ${\cal E}_g$, 
i.e. a map which transforms any input Gaussian state into a
Gaussian state. Other properties follow from those of the nG measures
for states.
\par
Despite the conceptual simplicity of the above definition, the evaluation
of $\delta[{\cal E}]$ is, in general, a challenging task using either the
HS-based or the QRE-based measure of nG. As a consequence, it has not
been used so far for a systematic classification of maps.
\section{non-Gaussianities of specific families of 
quantum states} \label{s:zoology}
This section is devoted to a sort of zoology of non-Gaussianity, i.e. we will 
consider different families of relevant quantum states and evaluate 
their non-Gaussianities $\delta_A$ and $\delta_B$. In this way, we 
analyze in some details the relationships between the two measures 
and illustrate their basic features also in connection with the 
analytical properties of their density operators and the intuition 
coming from their phase-space quasi-distributions.
\subsection{Fock states and superpositions}
\begin{figure}[h!]
\includegraphics[width=0.92\columnwidth]{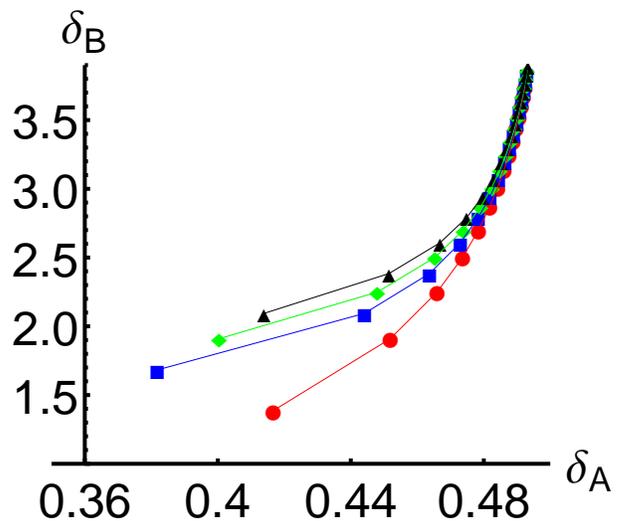}
\caption{
\label{f:3Fock}
QRE based nG $\delta_B$ as a function of HS distance nG 
$\delta_A$ for Fock states $|n\rangle$ with $n=1,\dots, 15$ 
(red circles) and for superpositions $|\psi_{nk}\rangle$ with 
$n=1,\dots 15$ and $k=3$ (blue squares), $k=4$ (green diamonds), and 
$k=5$ (black triangles).  }
\end{figure}

We consider single mode Fock states $|n\rangle$ and
superpositions of Fock states of the form 
$$|\psi_{nk}\rangle=\frac{1}{\sqrt{2}} \Big[|n\rangle 
+ | n+k \rangle\Big]\,,$$
for $n>0$ and $k >2$.
The reference Gaussian states are thermal states
$\tau_{nk}=\nu(n+\frac{k}2)$ with $n+\frac{k}2$ average photons. 
NG can be analytically evaluated for both measures
obtaining (for $k=0$ and $k>2$)
\begin{align}
\delta_A[\psi_{nk}] &= \frac{1}{2} \left(
1 + \frac{1}{2n+k} - 2 O_{nk} \right) \\ 
\delta_B[\psi_{nk}] &= h(n+\frac{k+1}2)\:,
\end{align}
where the overlap $O_{nk}=\langle\psi_{nk}| \nu (n+\frac{k}2) |
\psi_{nk}\rangle$ is given by
$$
O_{nk}= \frac12 \left[ 
\frac{(n+\frac{k}2)^{n}}{(n+\frac{k}2+1)^{1+n}}
+
\frac{(n+\frac{k}2)^{n+k}}{(n+\frac{k}2+1)^{1+n+k}}
\right]\,.
$$
As it is apparent from Fig. \ref{f:3Fock} both measures increases 
with both $n$ and $k$ and are monotone functions of each other 
for this families of states, \emph{i.e.} if 
$\delta_A[\varrho_1] > \delta_A[\varrho_2]$ then
$\delta_B[\varrho_1] > \delta_B[\varrho_2]$.
As stated in Lemma B7 Fock states have the maximum nG 
at fixed number of photons according to the measure $\delta_B$.
Though it has not been proved yet, we observe the same
result for the  HS-based $\delta_A$ in all the examples
considered up to now. 
\par
\subsection{Mixtures of Fock states}
\begin{figure}[h!]
\includegraphics[width=0.48\columnwidth]{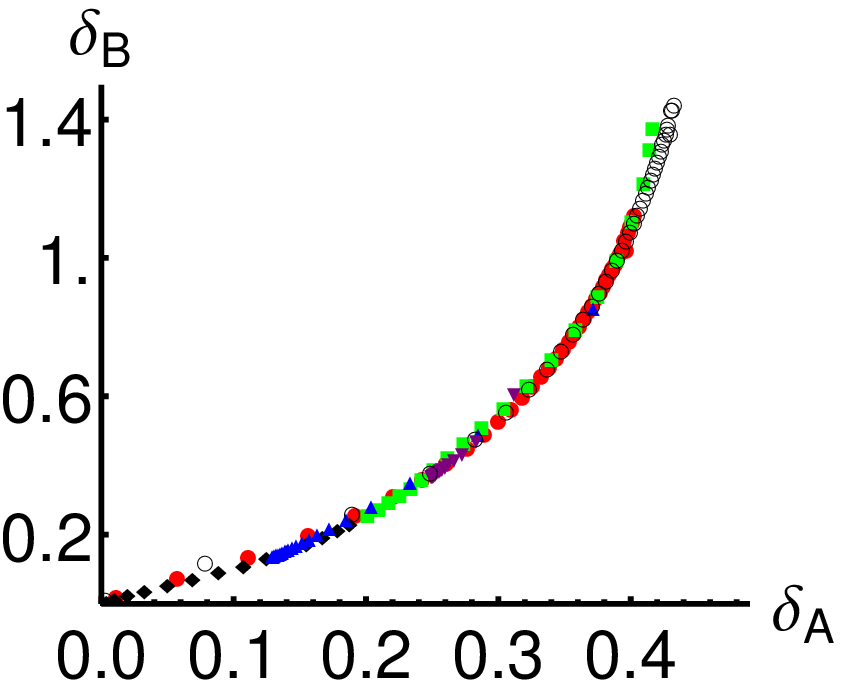}
\includegraphics[width=0.48\columnwidth]{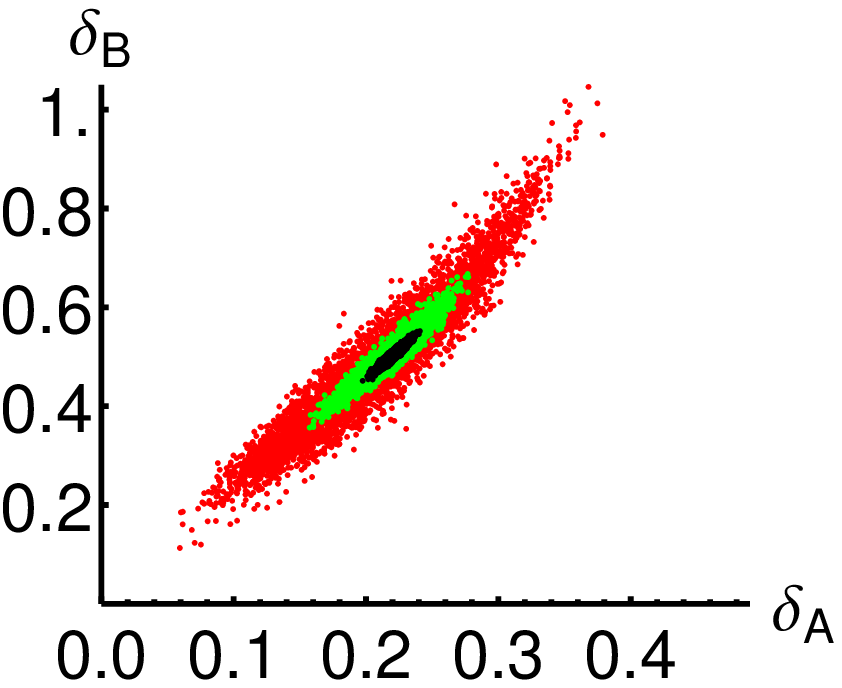}
\caption{(Color online)
\label{f:4mix}
Left panel: QRE based nG $\delta_B$ as a function of HS distance nG 
$\delta_A$ for mixtures of Fock states $\varrho_D$. Red circles
refer to a Poissonian distribution, whereas empty black circles are for
the diagonal states obitained as partial trace of TMC; green squares
and black diamonds are for mixtures coming from partial traces of 
photon-added and photon-subtracted two-mode squeezed vacuum
respectively; blue and purple triangles correspond to
$\Gamma$-distributions 
$q_n^{(2)}(\lambda)$ and $q_n^{(4)}(\lambda)$ respectively.
Right panel: 
QRE based nG $\delta_B$ as a function of HS distance nG 
$\delta_A$ for a sample of $10^4$ truncated random mixtures 
$\varrho_H$ of Fock states. The red cloud refers to 
$H=10$, green for $H=100$ and black for $H=1000$.}
\end{figure}

We now investigate the monotonicity of the
two measures for other one-parameter families
of quantum states. In this case we consider mixtures
of Fock states of the form
\begin{align}
\varrho_D = \sum_{n=0}^\infty q_n(\lambda)\,|n\rangle\langle n|
\end{align} 
where $0\leq q_n (\lambda) \leq 1$, $\sum_n q_n (\lambda)=1$ and  
$\bar n_\lambda =\sum_n n\,  q_n (\lambda)$ is the average photon 
number of the state. 
The reference Gaussian state for any diagonal mixtures of Fock 
states is a thermal state $\nu(\bar n_\lambda)$ with the 
same average photon number.  The non-Gaussianity can be thus 
written as
\begin{align}
\delta_A[\varrho_D] &= \frac12 \left[ 1
- \frac{\sum_n \tau_n(2 q_n - \tau_n) }{\sum_n q_n^2}\right] \notag \\ 
\delta_B[\varrho_D] &= h(\bar n_\lambda+1/2) + \sum_n q_n \log q_n\,,
\label{ngdiag}
\end{align}
where $\tau_n = \langle n|\nu(\bar n_\lambda)|n\rangle$ are the matrix elements
of the (thermal) Gaussian reference state.
We have numerically evaluated the non-Gaussianities for several one-parameter 
families including the diagonal states obtained as partial traces of 
TMS, PSSV and PASV states (see Eq. (\ref{eq:Poiss}) and the discussion following
Lemma B4) as well as diagonal states with Poissonian profile or given by
a $\Gamma$-distribution of the 
form $q_n^{(k)}(\lambda) \propto n^k \exp\{-n/\lambda\}$.
Results are shown in the left panel of Fig. \ref{f:4mix}: the two
measures are monotone each other for all the considered families and
the behaviour of $\delta_B[\varrho_D]$ vs. $\delta_A[\varrho_D]$ is 
almost independent on the kind of states. These results suggest that 
a general relation between the two measures may exist
for mixtures of Fock states. However, so far we have not been able to prove
it analytically starting from the expressions of $\delta_A[\varrho_D]$ and 
$\delta_B[\varrho_D]$ in Eqs. (\ref{ngdiag}).
\par
We have also considered (truncated) random mixtures of the 
form $$ \varrho_H = \sum_{n=0}^H p_n\, |n\rangle\langle n|\:,$$
where $H$ is the truncation dimension, and have numerically evaluated the
non-Gaussianities using Eqs. (\ref{ngdiag}) for a sample of $10^4$ 
states. Results are reported in the right panel of Fig. \ref{f:4mix} and 
show that despite the large number of involved parameters (up to 
$H=1000$) the two measures are almost monotone each other. 
As $H$ increases the distribution of the two measures concentrates
around the typical values.
\subsection{Schr\"{o}dinger cat states}
\begin{figure}[h!]
\includegraphics[width=0.48\columnwidth]{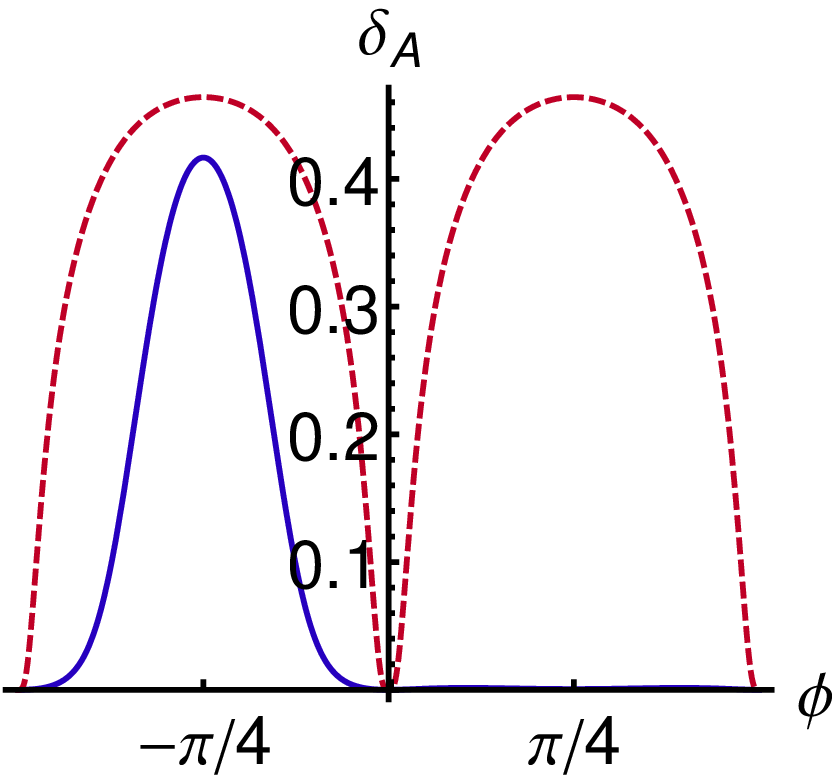}
\includegraphics[width=0.48\columnwidth]{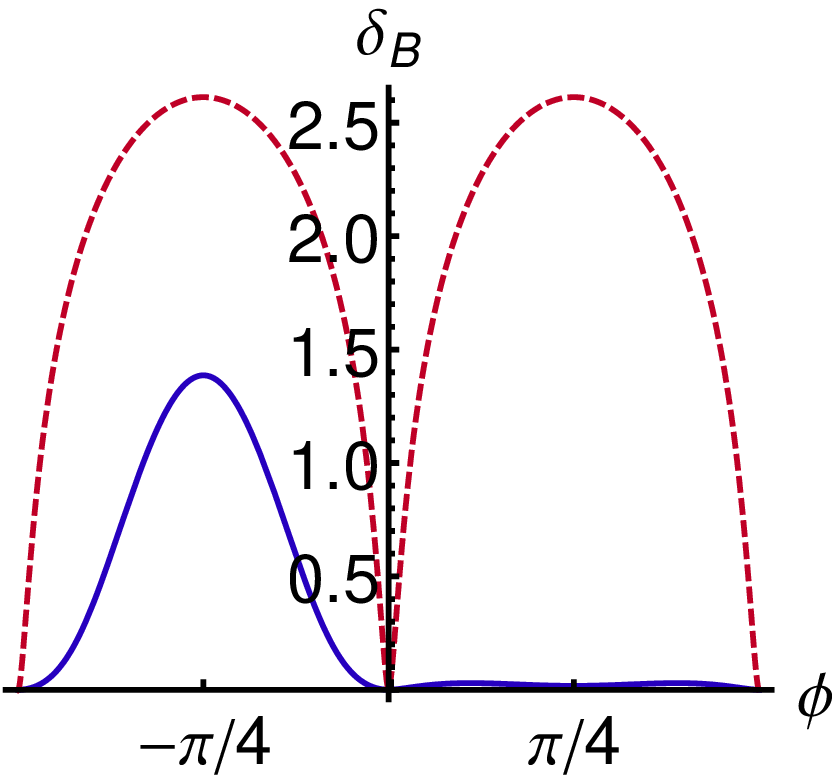}
\includegraphics[width=0.48\columnwidth]{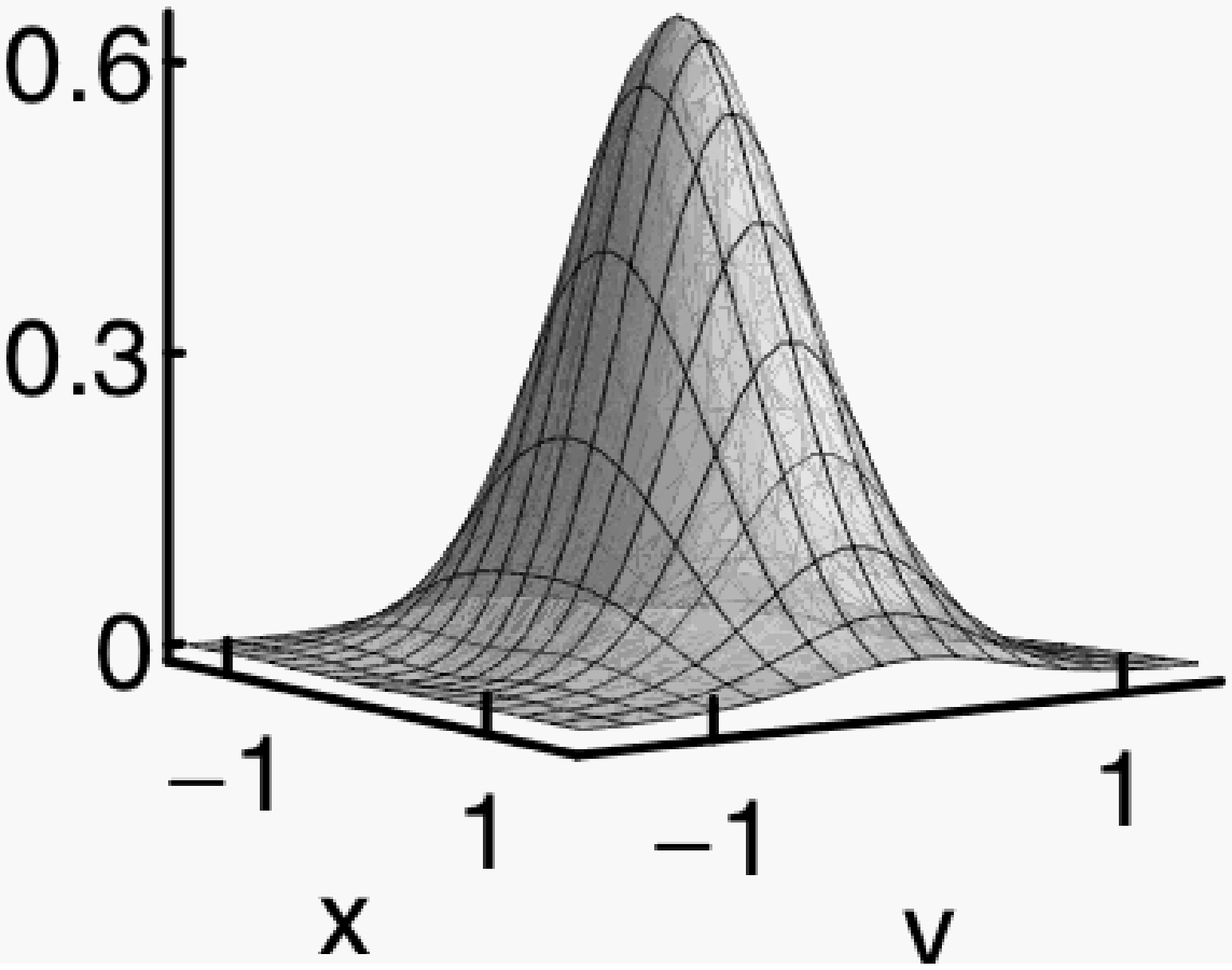}
\includegraphics[width=0.48\columnwidth]{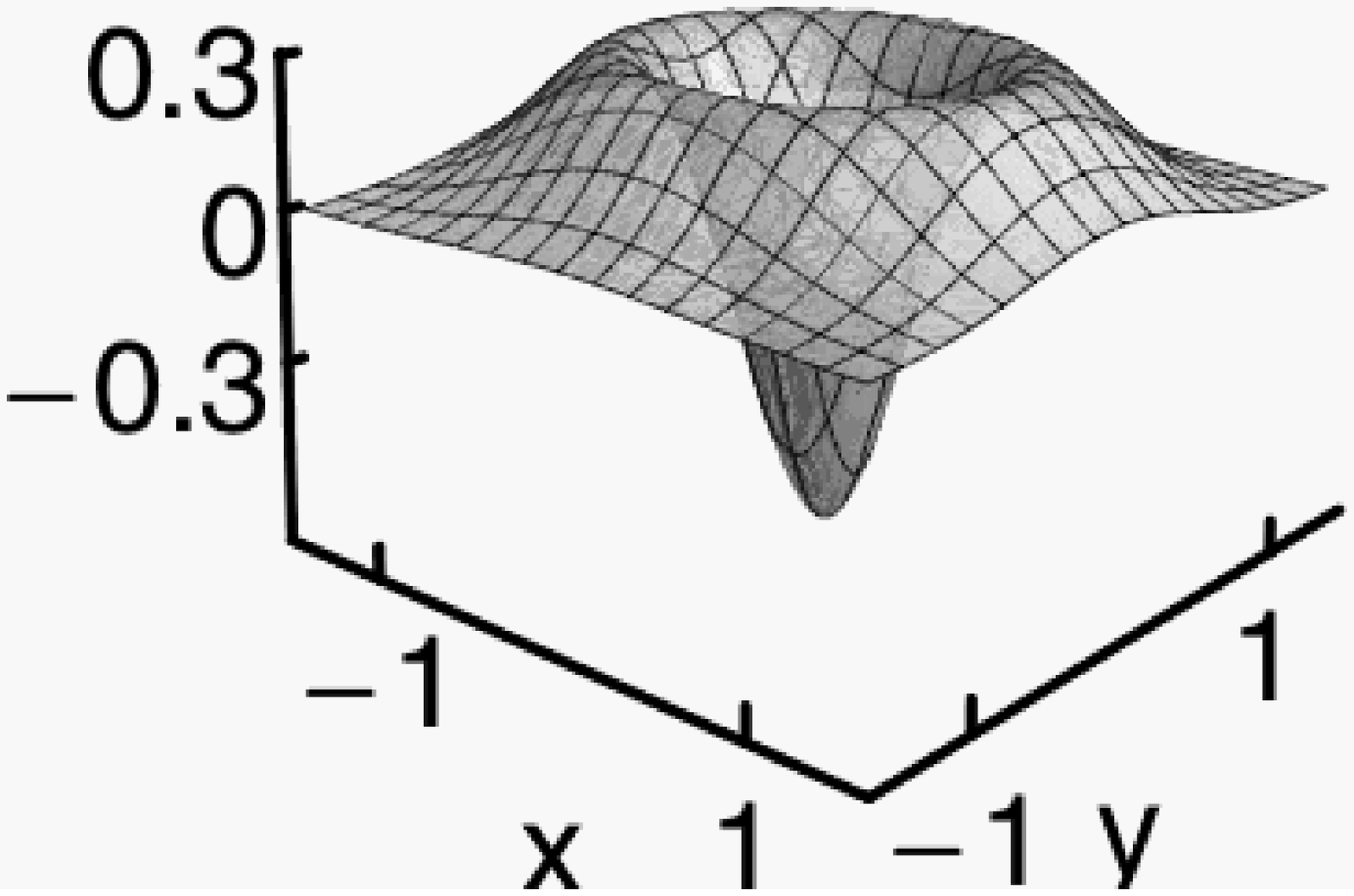}
\includegraphics[width=0.48\columnwidth]{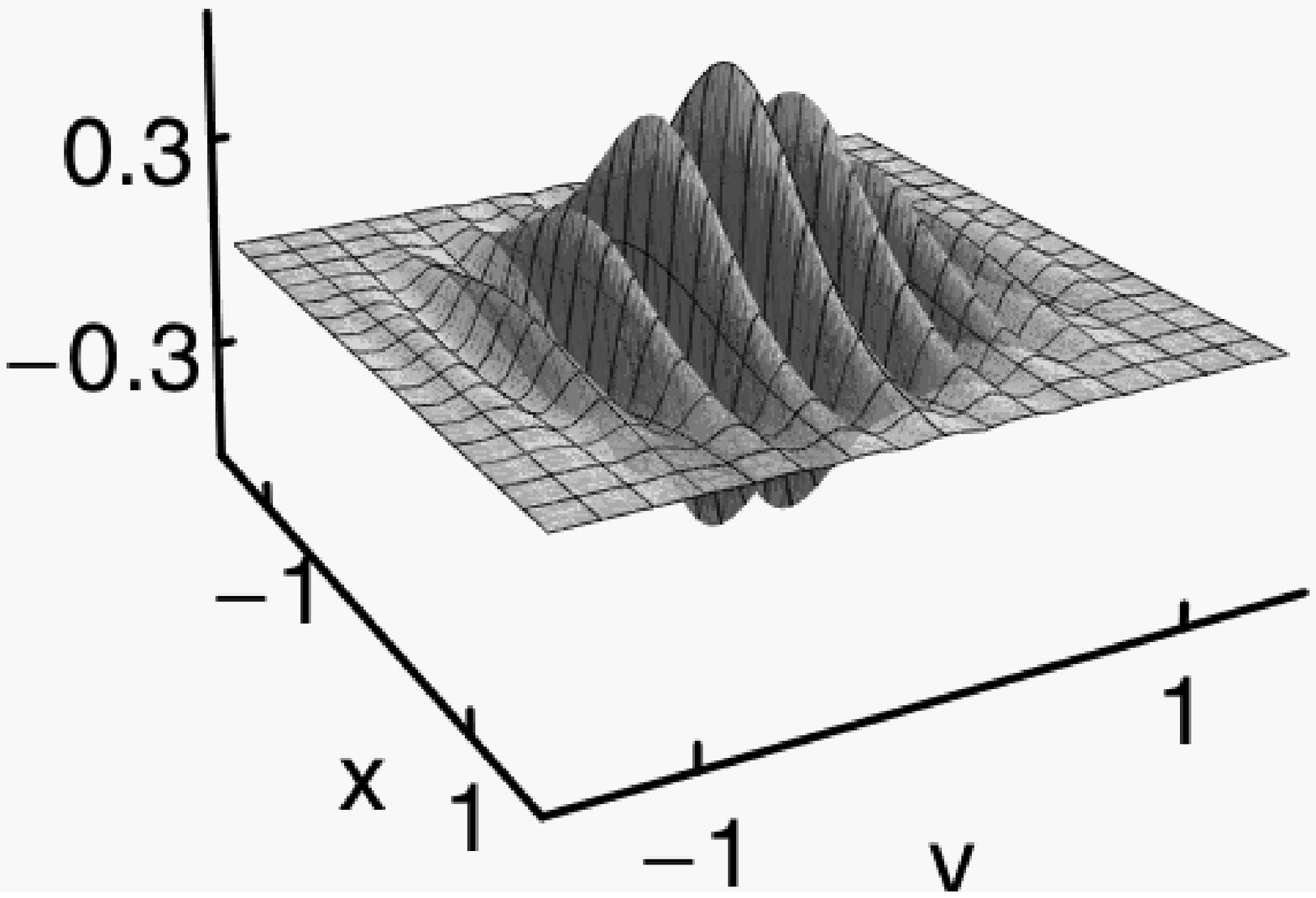}
\includegraphics[width=0.48\columnwidth]{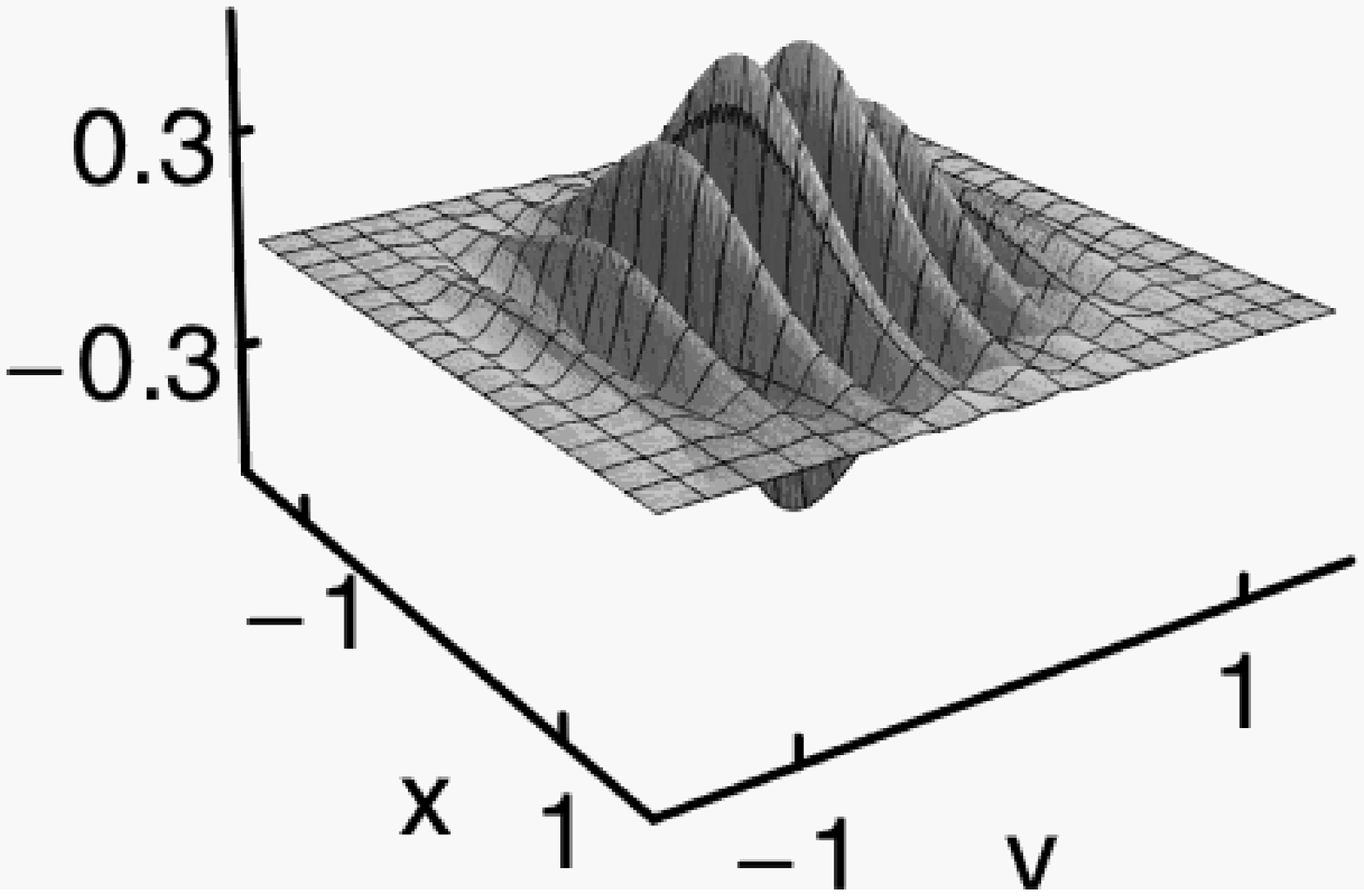}
\caption{
\label{f:5Cats}
HS based nG $\delta_A$ (left upper panel) and QRE based nG 
$\delta_B$ (right upper right) for  Schr\"{o}dinger  cat states 
$|\psi_S\rangle$ as a function of $\phi$ and for different
values of the amplitude $\alpha$. Solid bue line: $\alpha=
0.5$. Red dashed line: $\alpha=5$. The corresponding 
Wigner functions for the odd and the even cats are also
shown in the lower panels: 
$\alpha=0.5$, $\phi=+ \pi/4$ (top left); 
$\alpha=0.5$, $\phi=- \pi/4$ (top right);
$\alpha=5$, $\phi=+ \pi/4$  (bottom left);
$\alpha=5$, $\phi=- \pi/4$ (bottom right).
}
\end{figure}

Let us now consider the two-parameter family of quantum 
states given by the \emph{Schr\"{o}dinger  cat-like states}, that is
superpositions of coherent states 
$|\alpha\rangle = D(\alpha)|0\rangle$ and $|-\alpha\rangle$, 
\begin{align}
|\psi_S\rangle =  \frac{
\cos\phi |\alpha\rangle + \sin \phi |-\alpha\rangle }
{\sqrt{ 1+\sin(2\phi) \exp\{-2\alpha^2\} } }\:.
\end{align}
For $\phi=\pm \pi/4$, $|\psi_S\rangle$ reduces to the so-called 
odd and even Schr\"{o}dinger  cat states. Using the fact that
the reference Gaussian state
is a displaced squeezed thermal state $\tau_S=
D(C) S(r) \nu(N) S^{\dag}(r) D^{\dag}(C)$, where the
real parameters $C$, $r$ and $N$ are analytical function
of $\phi$ and $\alpha$ we have evaluated  
the nG measures $\delta_A$ and $\delta_B$ for different 
values of $\alpha$ and $\phi$. 
Results are shown in Fig. \ref{f:5Cats}. 
\par
As it is apparent from the plots, upon varying the value of
the parameters the two measures exhibit similar qualitative
behaviour. In particular, for low values of the amplitude 
(\emph{e.g.} $\alpha=0.5$), we observe an asymmetric 
behaviour with respect to $\phi$: nG is almost zero for positive
$\phi$, while for $\phi<0$ one achieve high values of nG.
By increasing the value of the amplitude, say $\alpha=5$, 
both measures become even function of $\phi$. This can be 
understood by looking at the Wigner functions of even and 
odd Schr\"{o}dinger  cat states. In fact for low amplitudes, 
the even cat ($\phi=\pi/4$) Wigner function
 is similar to a Gaussian state, 
in particular to the vacuum state, while for $\phi=-\pi/4$
it presents a non-Gaussian hole in the origin of the phase
space; for higher values of $\alpha$ one can observe similar
 non-Gaussian beahviours both for the even and the odd cat
state. 
\par
\begin{figure}[h!]
\includegraphics[width=0.94\columnwidth]{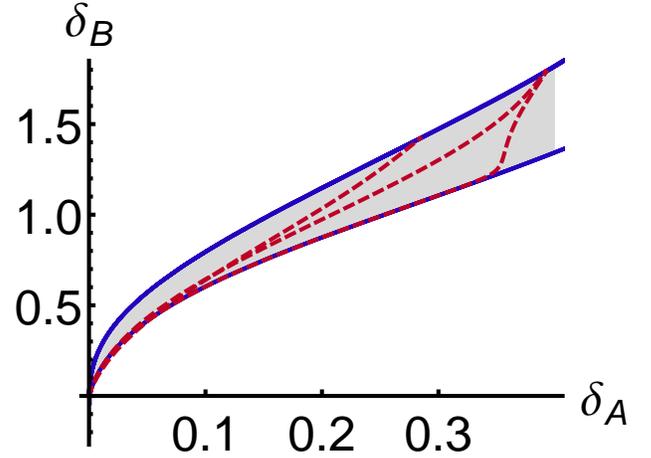}
\caption{
\label{f:6Cats}
QRE based nG $\delta_B$ as a function of HS distance nG 
$\delta_A$ for Schr\"{o}dinger  cat states $|\psi_S\rangle$. Solid 
blue lines refer to fixed amplitude of coherent states $\alpha$,
varying the angle $-\pi/2 < \phi < \pi/2$; from bottom to top we have
 $\alpha= 0.5, 2.5$. The dashed red lines are for fixed angles
$\phi$ and varying the amplitude $0<\alpha<2.5$; from bottom to top 
$\phi = -\pi/3, \pi/6, 2\pi/5$. The gray area denotes
all the allowed values for the two nG measures for the 
considered range of values of the two parameters.}
\end{figure}

Although the two nG measures capture the same qualitative 
non-Gaussian behaviour, it is apparent from the parametric 
plot of Fig. \ref{f:6Cats} that they induce different ordering on 
the set of states. In fact, upon varying the two parameters $\alpha$ and
$\phi$ one may find pair of states $\varrho_1$, $\varrho_2$ for which 
$\delta_A[\varrho_1] > \delta_A[\varrho_2]$ and $\delta_B[\varrho_1] 
< \delta_B[\varrho_2]$. As discussed before, we do accept that 
the two measures may induce different ordering on quantum states. 
Notice, however, that upon fixing one of the parameters and 
varying the other, one observes again a monotonous behaviour: 
the red dashed lines in Fig. \ref{f:6Cats} show $\delta_B$ vs 
$\delta_A$ for fixed values of $\phi$ and varying the amplitude 
$\alpha$. This appears to be a typical behaviour: 
for all the one-parameter families considered, the two measures 
are monotone with respect to each other and induce the same 
ordering of nG.
\section{Gaussification and de-Gaussification
processes} \label{s:GaussDeGauss}
In this section we will consider a single-mode Gaussification, the
\emph{loss} mechanism due to the interaction with a bath of harmonic
oscillators at zero temperature, and a de-Gaussification processes due
either to phase diffusion or Kerr interaction. 
Other Gaussification and de-Gaussification
protocols are considered in the next section where we discuss the case
of entanglement distillation.
\subsection{Loss mechanism}
The evolution of a single-mode quantum state interacting with
a bath of harmonic oscillators at zero temperature is described
by the following Lindblad Master equation
\begin{align}
\dot{\varrho}=\frac{\gamma}{2}\mathcal L[a]\varrho 
\end{align}
where $\dot{\varrho}$ denotes time derivative, $\gamma$ is the
damping factor and the superoperator $\mathcal{L}[O]$ acts as follows
\begin{align}
\qquad 
\mathcal L[O]\varrho= 2O^\dagger \varrho O -O^\dagger
O\varrho-\varrho O^\dagger O. \label{eq:Lindblad}
\end{align} 
Upon writing $\eta = e^{- \gamma
t}$ the solution of the Master equation can be written as
\begin{align}
\varrho(\eta) & = \sum_m V_m\:\varrho\: V_m^\dag \\
V_m & = \sqrt{\frac{(1-\eta)^m}{m!}}\, a^m\,\eta^{\frac{1}{2}(a^{\dag}a-m)}\,,
\nonumber
\end{align}
where $\varrho$ is the initial state. In particular 
if the system is initially prepared in a Fock state
$\varrho_p = |p\rangle\langle p|$, after the evolution
we obtain the mixed state
\begin{align}
\varrho_p(\eta) &= \sum_m V_m \varrho_p V_m^{\dag} =
\sum_{l=0}^{p} \alpha_{l,p}(\eta) |l\rangle\langle l| \label{eq:loss}
\end{align}
with 
\begin{equation}
\alpha_{l,p}(\eta) = \binom{p}{l} (1-\eta)^{p-l} \eta^l . \label{eq:alphaloss}
\end{equation}
Since the state is diagonal in the Fock basis, the reference
Gaussian state is a thermal state $\tau_p(\eta) = \nu (p\eta)$ 
with average photon number $p\eta$. Non-Gaussianity
$\delta_A$ can be evaluated analytically
\begin{align}
\delta_A &=  \frac{1}{2(1 - \eta)^{2m}\: {}_2 F_1\left(-m,-m,1;\frac{\eta^2}
{(\eta -1)^2}\right)} \nonumber \\
&\times \left\{(1 - \eta)^{2m}\: {}_2 F_1\left(-m,-m,1;\frac{\eta^2}
{(\eta -1)^2}\right) \right. \nonumber \\
&\left.  + \: (1 + 2m\eta)^{-1} - \frac{2(1 + (m-1)\eta)^m}
{(1+m\eta)^{m+1}} \right\}, \nonumber 
\end{align}
where ${}_2 F_1(a,b,c,;x)$ denotes the Hypergeometric function, and 
 while $\delta_B$
can be evaluated numerically via the formula
\begin{align}
\delta_B &= p\eta \log \left( \frac{p\eta + 1}{p\eta} \right) 
+ \log(1+p\eta)  \nonumber \\
& + \sum_{l=0}^\infty \alpha_{l,p}(\eta) \log ( \alpha_{l,p}(\eta) ). \nonumber
\end{align}
Because of Lemma B6 we know for sure that $\delta_B$ 
is decreasing with time $\eta t$,  
while this property is not guaranteed for $\delta_A$. 
\begin{figure}[h!]
\includegraphics[width=0.48\columnwidth]{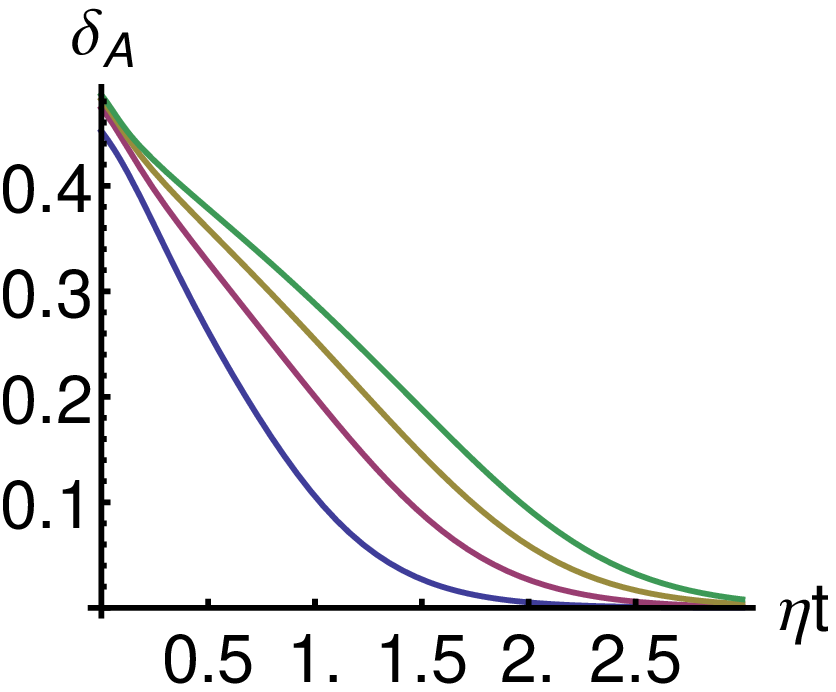}
\includegraphics[width=0.48\columnwidth]{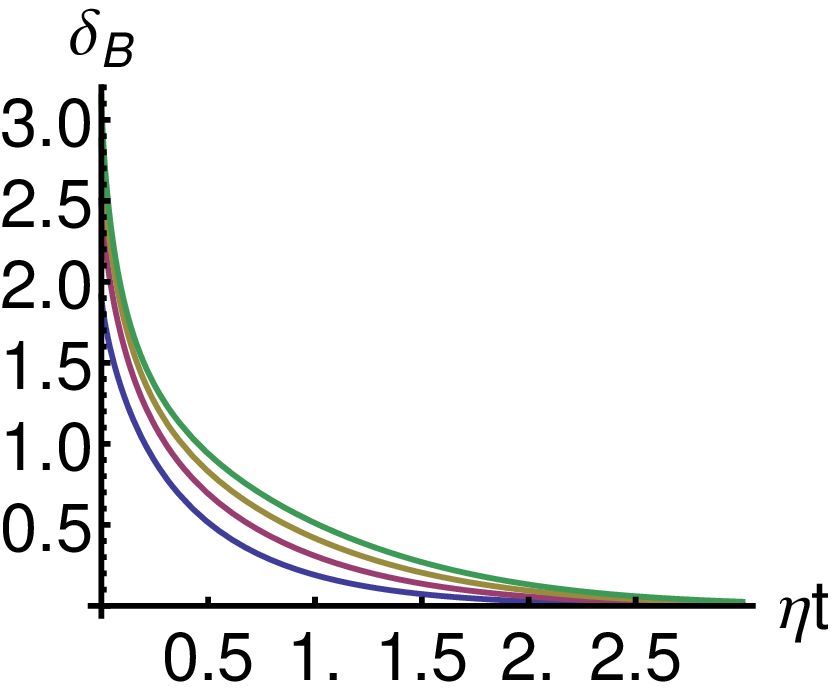}
\caption{
\label{f:7loss}
HS based nG $\delta_A$ (left panel) and QRE based nG 
$\delta_B$ (right panel) for a Fock state $|p\rangle$
under loss mechanism as a function of dimensionless time
$\eta t$ and for different values of $p$. From bottom to top
$p = \{ 2, 4, 6, 8 \}$.}
\end{figure} 
In Fig. \ref{f:7loss} we plot both non-Gaussianities as a function
of dimensionless time $\eta t$ and for different values of $p$, 
observing that also 
$\delta_A$ is decreasing with time, and that both are monotonically
increasing function of $p$, that is, at fixed time $t$ the higher 
is the initial photon number $p$, the larger is the nG of the 
evolved state. Although they present a different shape, due also
to their different scale ($\delta_A$ for single-mode states
is bounded by $1/2$, while $\delta_B$ is in general unbounded), 
we observe a similar trend for both nGs
which in particular approach zero for the same values of the 
parameters $\eta$, $t$ and $p$.
\subsection{Phase-diffusion evolution}
Let us consider single-mode systems evolving according to the following
Master equation
\begin{equation}\label{eq:PhDiffME} 
\dot\varrho= \Gamma\, \mathcal{L}[a^\dag a]\varrho,
\end{equation}
where the super-operator $\mathcal{L}[A]$ has been defined in 
Eq. (\ref{eq:Lindblad}) and $\Gamma$ is the noise factor.
This Master equation describes the evolution of a quantum state
subjected to a phase-diffusive noise. These non-Gaussian fluctuations 
are an important source of noise in optical communication links,
and protocols able to purify squeezing or distilling entanglement
have been recently proposed \cite{PurFiurPRL,EDHageNatPh}. 
Upon writing $\varrho$ in the Fock state basis, the ME leads 
to differential equations for the matrix elements 
$\varrho_{nm} = \langle n |\varrho |m\rangle$, where
$\dot\varrho_{nm} = -\frac12\, \Gamma\: (n-m)^2 \varrho_{nm}$
whose solutions read as follows
\begin{equation}\label{eq:PhDiffsol}
\varrho_{nm}(t) = e^{- \Delta^2(n-m)^2}\varrho_{nm}(0).
\end{equation}
In the last equation we defined $\Delta^2 \equiv \Gamma t /2$ whereas
$\varrho_{nm}(0)$ denote the matrix elements of the initial state.
>From Eq.~(\ref{eq:PhDiffsol}) it is clear that the off-diagonal 
elements of the density matrix are progressively destroyed, whereas the 
diagonal ones are left unchanged and, in turn, energy is conserved. 
\par
It is worth noticing that the same evolution as in 
(\ref{eq:PhDiffsol}) can be also obtained by the application 
of a random, zero-mean Gaussian-distributed phase-shift to the 
quantum state.  Since the phase shift of an amount $\varphi$ is 
described by the unitary operator $U_\varphi \equiv \exp(-i \varphi\, 
a^\dag a)$, we can write the state
degraded by the Gaussian phase noise as follows:
\begin{align}
\varrho_{\rm Gn} &= \int_{\mathbbm R}\!\!\! d\varphi\,
\frac{e^{-\varphi^2/(4 \Delta^2)}}{\sqrt{4 \pi \Delta^2}}\,
U_\varphi \varrho(0) U_\varphi^{\dag} \label{g:noise}\\
&= \sum_{nm}\int_{\mathbbm R}\!\!\! d\varphi\,
\frac{e^{-\varphi^2/(4 \Delta^2)}}{\sqrt{4 \pi \Delta^2}}
e^{-i\varphi (n-m)}\, \varrho_{nm}(0)  |n\rangle\langle m| \\
&= \sum_{nm} e^{-\Delta^2 (n-m)^2}\,\varrho_{nm}(0) |n\rangle\langle
m|\,.
\end{align}
The parameter $\Delta$ is related to the width
of the Gaussian distribution of the random phase-shift 
by the relation $\sigmaCM_{rnd}^2 = 2\Delta^2$:  as one may expect, 
the more the Gaussian distribution is broad, the higher is the phase-noise 
affecting the quantum state.  
\par
If we consider the mode initially prepared in a coherent state 
$|\alpha\rangle$ with real amplitude the phase-diffused state is
a non-Gaussian mixed state with density operator given by 
\begin{align}
\varrho_{\Delta,\alpha} = e^{-|\alpha|^2} \sum_{n,m}^{\infty} \frac{\alpha^{n+m} 
\:e^{-\Delta^2 (n-m)^2}}{\sqrt{n! \:m!}} |n\rangle\langle m|\,,
\end{align}
in the Fock basis.
\begin{figure}[h!]
\includegraphics[width=0.48\columnwidth]{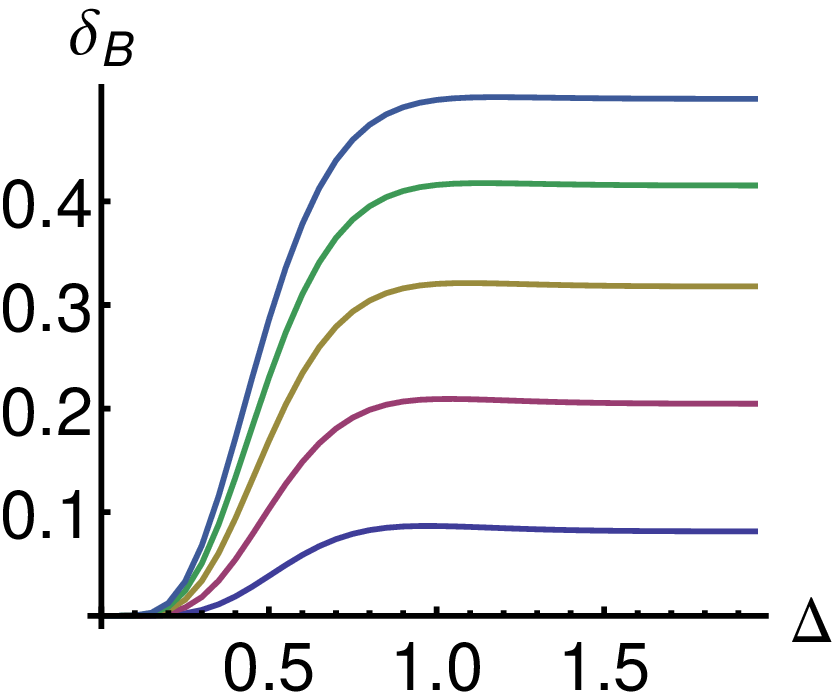}
\includegraphics[width=0.48\columnwidth]{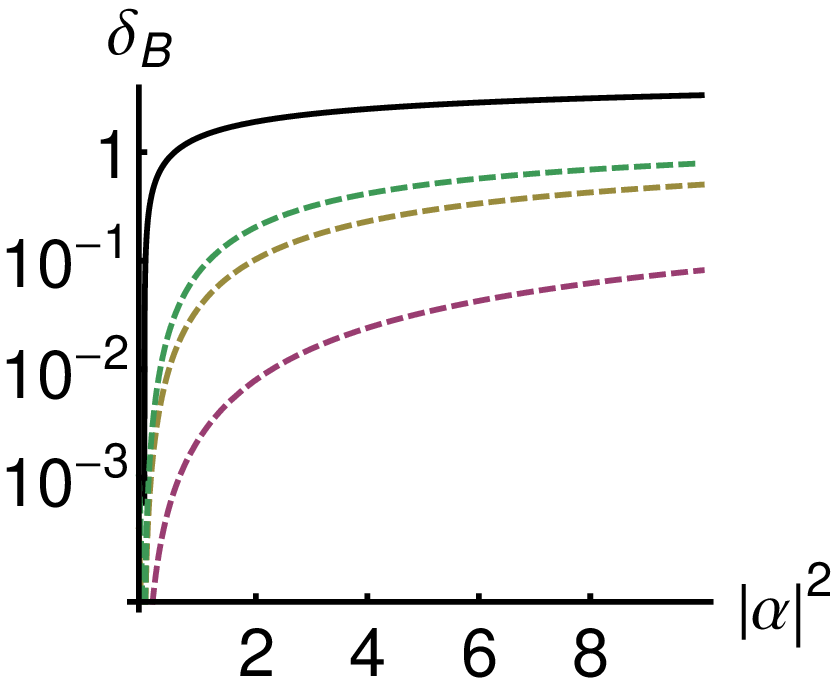}
\caption{
\label{f:078PhDiff}
(Left panel): QRE based nG $\delta_B$ for coherent states
undergoing phase-diffusion as a function of the
noise parameter $\Delta$ and for different values of the average
number of photons. From bottom to top: $|\alpha|^2 = \{ 1, 2, 3, 4, 5\}$.
(Right panel): QRE based nG $\delta_B$ for coherent states
undergoing phase-diffusion as a function of the
average number of photons $|\alpha|^2$ for different values noise
parameter: dashed lines, from bottom to top: $\Delta= \{ 0.25,0.5, \infty  \}$. 
The black solid line represent the maximum value of nG at fixed 
number of photons.}
\end{figure} 
\par
We have evaluated numerically the nG $\delta_B[\varrho_{\Delta,\alpha}]$ 
for different values of the noise parameter $\Delta$ and of the 
average number of photons $|\alpha|^2$. The results are shown in 
Fig. \ref{f:078PhDiff}. In the left panel we report $\delta_B$ 
as a function of $\Delta$ and for different values of $\alpha$: 
large values of nG are achieved and the more intense is the initial coherent state, 
the more non-Gaussian is the output. For large values of the noise
parameter the off-diagonal elements of the density matrix are completely
destroyed and the nG approaches its aysmptotic value, corresponding
to the nG of a diagonal mixture of Fock states with a Poissonian
distribution. The asymptotic value may be evaluated using Eq. (\ref{ngdiag}) 
and choosing  $q_n$ as a Poissonian distribution with mean value 
$\lambda=|\alpha|^2$. Before approaching the asymptotic value 
$\delta_B[\varrho_{\Delta,\alpha}]$ is not monotone: a not much 
pronounced maximum of $\delta_B$ may be seen for intermediate values
of $\Delta$. 
In the right panel of Fig. \ref{f:078PhDiff} we plot the nG for different
values of $\Delta$ as a function of the number of photons $|\alpha|^2$.
As we noticed above the maximum value of 
$\delta_B[\varrho_{\Delta,\alpha}]$ at fixed $|\alpha|^2$
is not the asymptotic value. However, the difference is very small and
thus, upon observing  the behaviour for different values of $\Delta$, we 
conclude that the maximum nG cannot be achieved 
by this family of quantum states.
\subsection{Kerr interaction}
\begin{figure}[h!]
\includegraphics[width=0.9\columnwidth]{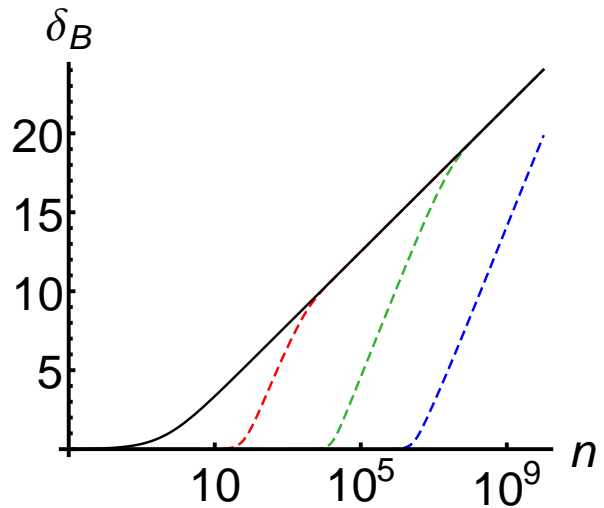}
\caption{
\label{f:8Kerr}
QRE based nG $\delta_B$ of coherent states undergoing Kerr interaction as
a function of the average number of photons and for different values of the
coupling  constant $\gamma$. Dashed lines from bottom to top $\gamma = \{ 10^{-6}, 
10^{-4}, 10^{-2} \}$.
The black solid line is the maximum nG at fixed number of photons.}
\end{figure}

One of the simplest unitary non-Gaussian evolution is provided by 
the so-called self-Kerr effect taking place in third-order nonlinear
$\chi^{(3)}$ media. The interaction Hamiltonian is given by
\begin{align}
H_{kerr} = \Gamma \,  (a^\dagger a )^2
\end{align}
and the evolution operator by $U_{kerr} = \exp \{
-i\gamma\:  (a^\dagger a )^2
\}$ where $\gamma = \Gamma t$ is a 
dimensionless coupling constant. Kerr interaction has been suggested to 
realize quantum nondemolition measurements, to 
enhance quantum estimation performances in quantum 
optics and to generate quantum superpositions 
\cite{KerrSup,KerrSup1} as well 
as squeezing \cite{Sun96} and entanglement 
\cite{kerr:PRL:01,kerr:PRA:06}. 
A known example of Kerr medium is provided by optical fibers where, however, 
nonlinearities are small and accompanied by other unwanted effects. 
Recently, larger Kerr nonlinearities have benn proposed in many different 
physical systems \cite{KerrProposals} and have been observed with electro-magnetically 
induced transparency \cite{KerrEIT}, with Bose-Einstein condensates \cite{KerrBEC} 
and with cold atoms \cite{KerrCA}. These results renewed the interest
for the quantum effects of Kerr interaction, which are always
accompanied by the generation of non-Gaussianity.  In the following,
we consider the nG features of an initial coherent state $|\alpha\rangle$ 
undergoing Kerr interaction
\begin{align}
|\alpha_\gamma\rangle =
 U_{kerr}|\alpha\rangle = e^{-|\alpha|^2/2}\sum_{n=0}^\infty
 \frac{\alpha^n}{\sqrt{n!}} e^{-i \gamma n^2} |n\rangle\:.
\end{align}
Since the evolution is unitary, the output state is still pure and 
the evaluation of the nG $\delta_B$ is straightforward 
and involves only the computation of the covariance matrix 
of the evolved state. Non-Gaussianity $\delta_B[|\alpha_\gamma\rangle ]$
is plotted in Fig. \ref{f:8Kerr} as a function of the
average number of photons $n=|\alpha^2|$ and for different
values of the coupling constant $\gamma$. The maximum nG achievable
for given number of photons is also reported for comparison. 
As expected, non-Gaussianity 
is an increasing function of the initial energy $n$ and, 
for the range of values here considered, of the the 
coupling constant $\gamma$.
\par
For $\gamma\approx 10^{-2}$ the maximum nG achievable at fixed energy
is quite rapidly achieved, while for more realistic values of the Kerr
coupling nG is obtained only for large values of the average
number of photons. In the experiments proposed to 
obtain entanglement via Kerr interaction \cite{kerr:PRL:01,kerr:PRA:06}, 
pulses with an average number up to $10^8$ photons are
needed to compensate the small nonlinearities of standard glass
fibers. Therefore, in these regimes, the generation of entanglement is always
accompanied by a large degree of nG.
\section{NG and distillation of
entanglement} \label{s:distillation}
Long distance quantum communication requires that the communicating
parties share highly entangled states over long distance. One has
therefore to deal with the daunting task of distributing highly
entangled states over long distances, overcoming losses and decoherence
due to the unavoidable coupling of the system with the environment. For
discrete-variable states (qudit) entanglement distillation protocols,
where a subset of states with an high degree of entanglement are
extracted from an enemble of less entangled states, have been proposed
and experimentally demonstrated.  As regards continuous-variable
entanglement, it has been proved that entanglement distillation cannot
be performed within the Gaussian world, i.e. by starting from Gaussian 
entangled states and by linear optical components, homodyne detection 
and classical communication. Non-Gaussianity is a necessary ingredient 
in an entanglement distillation protocol. In particular, two different 
main approaches can be adopted: in the first approach, e.g. the protocol
of \cite{DistBrowne,EDHageNatPh}, one starts with an entangled
non-Gaussian state, and then use Gaussian operations based on linear
optical elements, homodyne detection or vacuum projective measurements.
In the second approach one starts with an entangled Gaussian beam and
try to increase its entanglement by using non-Gaussian operations such as
photon number conditional measurements.  In
\cite{nGTLP1,nGTLP2,nGTLP3} it has been proved that two-mode
squeezed states with photons subtracted on the two modes can be used to
obtain better teleportation fidelities. A first full scheme of
entanglement distillation based on this idea has been presented in \cite{DistTaka}, 
where increase of entanglement by means of local photons subtraction from two-mode
Gaussian states has been observed.  In the following we will review 
the protocols of \cite{DistBrowne} and \cite{DistTaka} pointing out the role 
played by non-Gaussianity and its amount in the success of the protocols. 
\par
The protocol proposed in \cite{DistBrowne}, from now one the B-protocol, makes 
use of beam splitters and on/off detectors, i.e. detectors only able to 
distinguish the presence or the absence of photons.  The input state of 
the protocol is the state $\varrho\otimes\varrho$ i.e. two replicas of a 
two-mode non-Gaussian. The two copies are mixed in a balanced beam
splitter and then two of the output modes are directed into on/off photon
detectors. The state is kept if both the local detectors are registering the 
outcome "zero", i.e., at least in ideal conditions, the presence of the
vaccum state. In \cite{DistBrowne} it has been proved that the B-protocol
drives the initial state towards a zero-displacement Gaussian state,
which in turn are the only fixed points of the map. Moreover, under some 
assumptions on the initial state, the protocol acts as a purifying protocol 
as well, distilling the initial non-Gaussian entangled mixed state into
a pure Gaussian entangled state. Notice that despite the fact that the output 
state of each step is obtained via (Gaussian) passive linear operations
and Gaussian measurements (the projection on the vacuum state), the map cannot 
be described by a proper completely positive Gaussian operation 
\cite{noteGaussianMap}. 
\begin{figure}[h!]
\includegraphics[width=0.92\columnwidth]{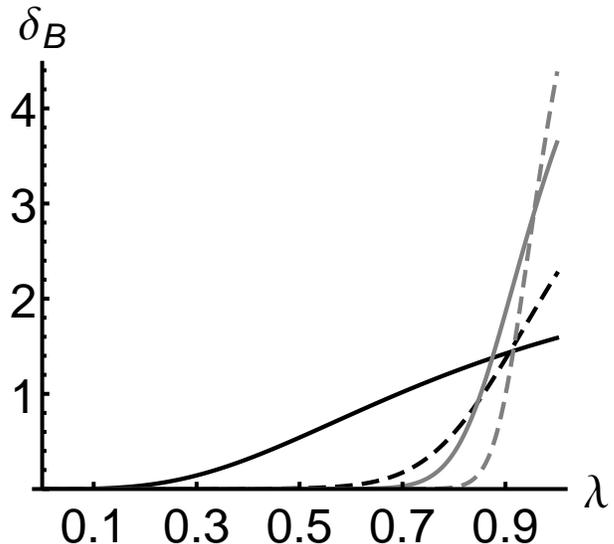}
\caption{Non-Gaussianity in the B-protocol. The plot shows the QRE based 
non-Gaussianity $\delta_B$ of the output state after $s$ steps of the B-protocol, 
when the initial state $\varrho_a$ is given in Eq.~(\ref{eq:browneA}), 
as a function of the parameter $\lambda$. 
Step: $s=0$ - black solid line;
$s=5$ - black dashed line; 
$s=10$ - gray solid line;
$s=20$ - gray dashed line.}
\label{f:9browne1}
\end{figure}
\par
As an illustrative example let us consider the B-protocol applied to 
the non-Gaussian pure state with the following non-zero matrix elements 
in the Fock basis 
$\varrho(a,b,c,d)=\langle a,b|\varrho|c,d\rangle$ 
\begin{align}
\varrho_a(0,0,0,0) &= \frac{1}{1+\lambda^2} \nonumber \\
\varrho_a(0,0,1,1) &= \varrho_a(1,1,0,0) = \frac{\lambda}{1+\lambda^2} 
\nonumber \\
\varrho_a(1,1,1,1) &= \frac{\lambda^2}{1+\lambda^2}\,. \label{eq:browneA}
\end{align}
In Fig. \ref{f:9browne1} we show the non-Gaussianity 
quantified by $\delta_B$ as a function of the parameter $\lambda$.
for states obtained after different number of steps of the protocol.
As a matter of fact, at fixed $\lambda$, nG is not always 
monotonically decreasing under the iteration of the protocol, 
and for $\lambda \approx 1$ the value of nG may increase, even 
achieving very high values. On the other hand, the
overall effectiveness of the protocol is confirmed by 
our measure, since the range of values of $\lambda$ for 
which $\delta_B\approx 0$ increases at each step of the 
protocol. Overall, the use of our nG measure may help to tailor the 
distillation protocol depending on the initial conditions.
\par
\begin{figure}[h!]
\includegraphics[width=0.48\columnwidth]{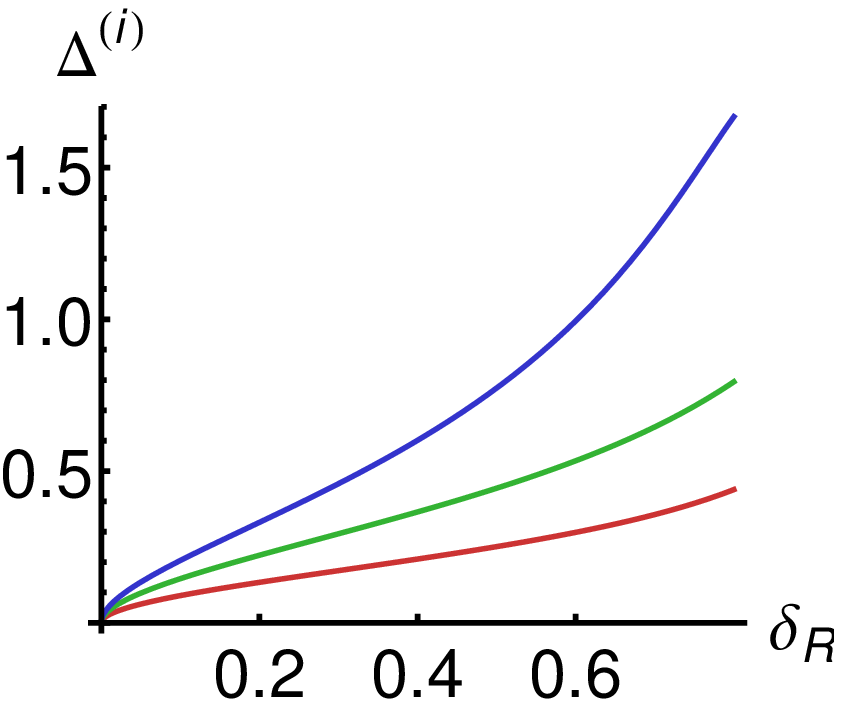}
\includegraphics[width=0.48\columnwidth]{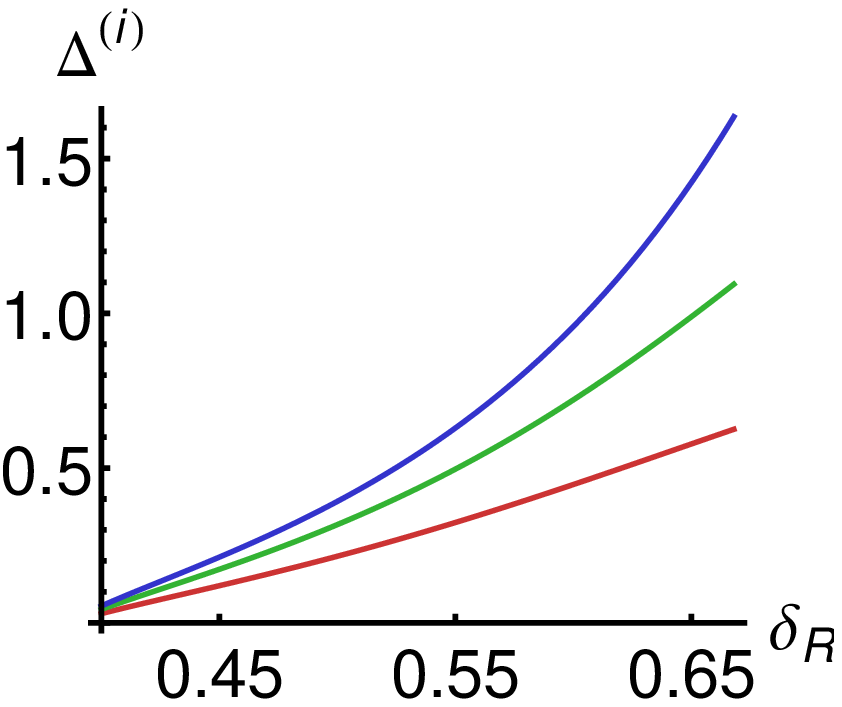}
\caption{The relative increase of entanglement $\Delta^{(i)}$ at the
$i$-th step of the B-protocol as a function of the renormalized nG 
of the initial state (left panel: $\varrho_A$; right panel:
$\varrho_B$). From bottom to top: $i = \{1, 2, 5, \infty \}$
\label{f:10browne2}}
\end{figure}

Let us now consider as initial states the pure state $\varrho_a$ of 
Eq. (\ref{eq:browneA}) and the mixed state $\varrho_b$ with 
non-zero matrix elements
\begin{align}
\varrho_b(0,0,0,0) &= \frac{1}{1+\lambda^2} \nonumber \\
\varrho_b(0,0,1,1) &= \varrho_b(1,1,0,0) = \frac{\lambda}{2(1+\lambda^2)} 
\nonumber \\
\varrho_b(1,1,1,1) &= \frac{\lambda^2}{1+\lambda^2}\,, \label{eq:browneB}
\end{align}
Both states converge towards pure Gaussian entangled states, and it has 
been shown \cite{DistBrowne} that entanglement increases at each step of 
the protocol. Here we will investigate how much the gained entanglement is 
related to the non-Gaussianity of the initial state. For this purpose, since 
both, entanglement and non-Gaussianity are increasing quantities with the number
of photons, we will consider a renormalized version of non-Gaussianity $\delta_R[\varrho]$ 
and the relative entanglement gain at each step $\Delta^{(i)}$.
The maximum amount of non-Gaussianity for a two-mode state
with $N=\Tr[\varrho (a^{\dag}a + b^{\dag}b)]$ photons is 
\begin{align}
\delta_M^{(2)} (N) = 2[ (1+ N/2) \log(1+ N/2) - (N/2) \log(N/2) ], \nonumber
\end{align}
and the renormalized non-Gaussianity is defined as
\begin{align}
\delta_R[\varrho] = \frac{\delta_B[\varrho]}{\delta_{M}^{(2)}(N)}.
\end{align}
We define the relative entanglement gain at the step $i$ as
\begin{align}
\Delta ^{(i)} = \frac{ E_N(\varrho^{(i)}) - 
E_N(\varrho^{(0)}) }{E_N(\varrho^{(0)})}
\end{align}
where $\varrho^{(i)}$ is the output state at the $i$-th
step of the protocol. The degree of entanglement is
quantified in terms of the logarithmic negativity, i.e.
\begin{align}
E_N(\varrho)
=\log_2 \lVert \varrho^\Gamma \rVert_1
\end{align}
where $\lVert \,\, \rVert_1$ denotes the trace-norm, and 
$\varrho^{\Gamma}$ is the partial transpose of $\varrho$
\cite{PPT-Peres}. In Fig. \ref{f:10browne2} we plot for both
states $\varrho_a$ and $\varrho_b$, 
the increase of entanglement $\Delta^{(i)}$ 
as a function of the renormalized non-Gaussianity $\delta_R[\varrho]$.
We observe that the more the initial state is non-Gaussian, the 
larger is the entanglement increase at each step of the protocol. 
Therefore, we may say that at least for this particular
protocol non-Gaussianity plays a relevant role, and it is quantitatively 
responsible for the good performances of the protocol.
\par
\begin{figure}[h!]
\includegraphics[width=0.48\columnwidth]{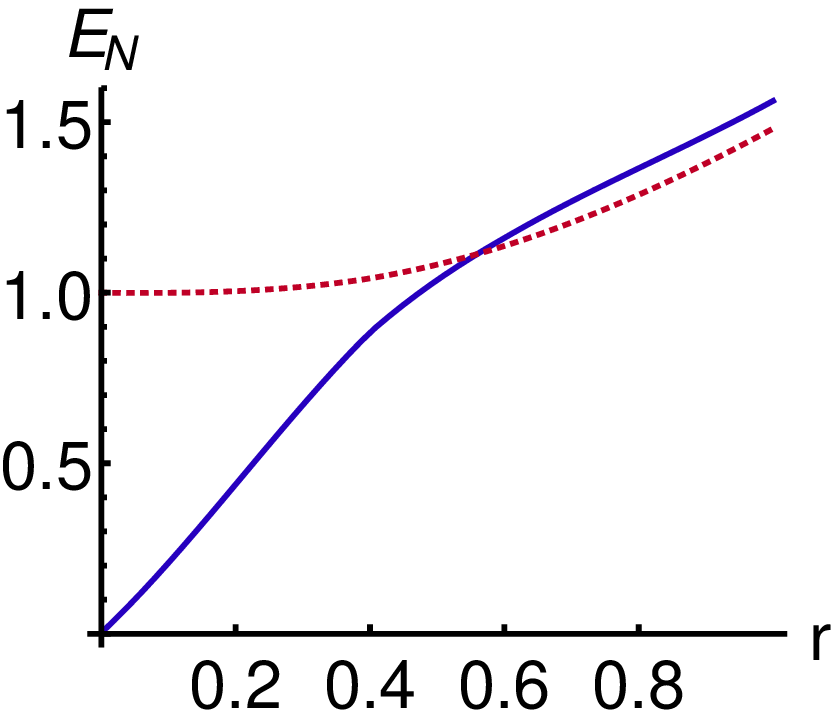}
\includegraphics[width=0.48\columnwidth]{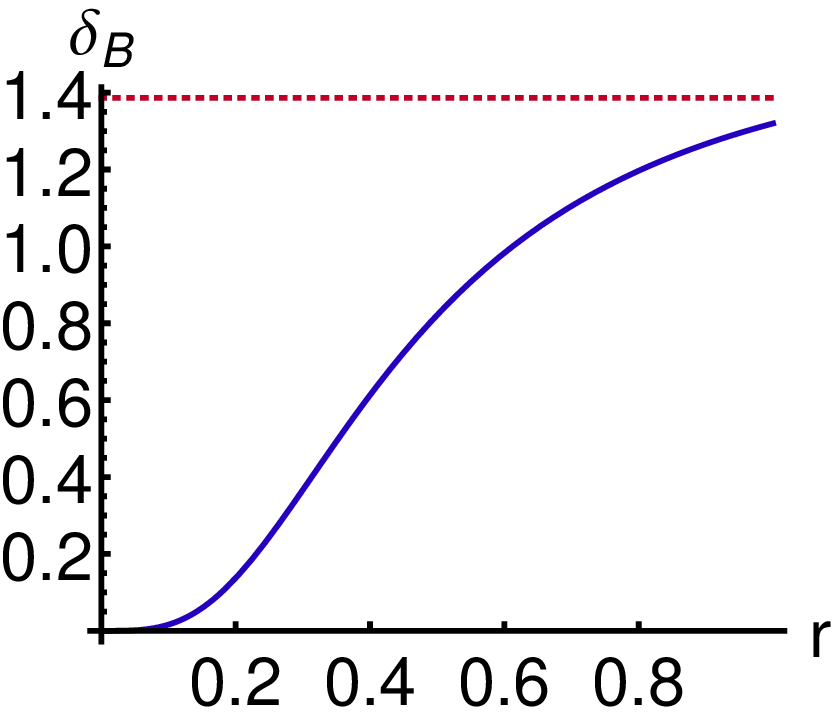}
\caption{Entanglement (left panel) and QRE based
non-Gaussianity (right panel) of the distilled states 
in the T-protocol as a function of the squeezing parameter $r$ . 
Dotted red: one-photon subtracted state $|\psi_{{\scriptstyle out}}^{(1)}\rangle$. 
Solid blue: two-photon subtracted state $|\psi_{{\scriptstyle out}}^{(2)}\rangle$.
\label{f:11taka}}
\end{figure}

Let us now consider the distillation protocol proposed in 
\cite{DistTaka}, from now on the T-protocol. In this protocol, an 
entangled Gaussian state is obtained by mixing a single-mode 
squeezed state $|\psi\rangle = S(r)|0\rangle$ with the vacuum in 
a balanced beam-splitter. Then, both the modes are splitted to perform
a photon-subtraction and the state is keeped if one or both
detectors clicks (see \cite{DistTaka} for details). 
The two possible output states of the distillation procedure 
can be written in a compact form as
\begin{align}
|\psi_{{\scriptstyle out}}^{(i)}\rangle_{AB}&=
\mathcal{N} \,a_A^{n_A}a_B^{n_B}\hat{B}_{AB}(\pi/4) 
\hat{S}_A(r) |0\rangle_A |0\rangle_B \\
 &=\mathcal{N} \hat{B}_{AB}(\pi/4)
 \,a_A^{n_A+n_B}\hat{S}_A(r)|0\rangle_A |0\rangle_B
\end{align}
where if $(n_A,n_B)=(1,0)=(0,1)$ we have a single-photon subtracted
state $|\psi_{{\scriptstyle out}}^{(1)}\rangle$ , while for $(n_A,n_B)=(1,1)$ we obtain
the two-photon subtracted state  $|\psi_{{\scriptstyle out}}^{(2)}\rangle$. 
The non-Gaussianities of the two states can be easily evaluated
by using the properties of the measure $\delta_B$, and in particular by
exploiting the invariance under Gaussian unitary operation such as
beam-splitter evolution and squeezing. After some calculations, it turns out that
\begin{align}
\delta_B[|\psi_{{\scriptstyle out}}^{(1)}\rangle] &= \delta_B[|1\rangle] \nonumber \\
\delta_B[|\psi_{{\scriptstyle out}}^{(2)}\rangle] &= 
\delta_B[\mathcal{N}^\prime(\mu |0\rangle + \sqrt{2}\nu|2\rangle)] \nonumber
\end{align}
where $\mu=\cosh(r)$, $\nu=\sinh(r)$  and $\mathcal{N}^\prime$ is a
normalization factor. We can immediately observe that for 
$|\psi_{{\scriptstyle out}}^{(1)}\rangle$
the nG does not depend on the squeezing parameter $r$.
In Fig.~\ref{f:11taka} we plot as a function of 
$r$ respectively in the left panel the entanglement and
in the right panel the non-Gaussianities of the two possible ouput 
states. We observe that for small values of $r$ the entanglement
and the nG of the output states have a similar beahviour. However
this similarity is lost when we consider higher values of the squeezing, 
i.e. we can say nothing in general about the relationship between 
the distilled entanglement and the non-Gaussianity for this protocol.
\par
In summary, the role of the amount of non-Gaussianity 
in continuous variable entanglement distillation protocols is still 
not  fully understood and worth to study for protocols employing
non-Gaussian entangled states as initial resource.
In particular, an interesting open question that arises in this framework is 
whether there is a maximum amount of distillable entanglement at fixed
non-Gaussianity of the resource state.
\section{NG and quantum communication} \label{s:entropic}
As anticipated in the introduction, Gaussian states play a relevant 
role in quantum  communication, in particular for being
\emph{extremal,} at fixed covariance matrix for 
several relevant quantities \cite{Wolf}. 
For example, channel capacities are maximized by Gaussian states, 
whereas most of the entanglement measures are minimised by Gaussian 
states. In this section we will prove the extremality conditions for 
some of these quantities by using our non-Gaussianity measure 
$\delta_B$, and studying the role played by non-Gaussianity
itself in these inequalities
\par
As a first example, let us consider a very important and well known 
quantity in quantum information, i.e. the $\chi$-Holevo quantity. Let 
us suppose to transmit information by encoding the symbols $a_i$, chosen
with probabilities $p_i$ from an alphabet ${\cal A}$, in a set of 
quantum states $\varrho_i$. At each use of channel, the state
preparation is described by the overall state $\varrho=\sum_i p_i \varrho_i$, 
and the Holevo quantity, which quantifies the amount of accessible 
information, is given by the formula
$\chi(\varrho) = \S(\varrho) -\sum_i p_i \S(\varrho_i)$. 
If we fix the covariance matrix of the overall state $\varrho$ 
and consider pure encoding states $\varrho_i$, we obtain
(we will denote $\delta_B = \delta $ in the rest of the paper),
\begin{align}
\chi(\varrho) = \S(\tau) - \delta[\varrho]
\end{align}
The maximum value of $\chi$ is attained by 
considering the Gaussian state and the nG 
$\delta[\varrho]$ exactly quantifies how much information is lost 
by considering a non-Gaussian overall state.
\par
Let us now consider bipartite states and rephrase the extremality condition
for two quantities: the quantum mutual information and the quantum
conditional entropy. The quantum mutual information, given a bipartite
state $\varrho_{AB}$ is defined as $I(A:B) = \S(\varrho_A) +
\S(\varrho_B) -\S(\varrho_{AB})$ \cite{Cer97}. It quantifies the amount of
correlations (classical and quantum) in a bipartite state and, for pure
states corresponds to the entanglement. In  \cite{QMIHolevo} it has been
shown that at fixed covariance matrix $I(A:B)$ is minimised by Gaussian
states. By using our measure $\delta$ the proof  is simply based on the
lemma B5, that is
\begin{align}
I(A:B) &= \S(\varrho_A) + \S(\varrho_B) -\S(\varrho_{AB}) \nonumber \\
&= \S(\tau_A) + \S(\tau_B)  -\S(\tau_{AB}) \nonumber \\
& \qquad + (\delta[\varrho_{AB}] - \delta[\varrho_A] 
-\delta[\varrho_B]) \nonumber \\
&\geq I_G (A:B) 
\end{align}
where $I_G(A:B) = \S(\tau_A) + \S(\tau_B)  -\S(\tau_{AB})$ is the 
quantum mutual information obtained by considering the 
reference Gaussian states. The two mutual informations $I(A:B)$
and $I_G(A:B)$
differs exactly by the quantity 
$$\Delta_2= \delta[\varrho_{AB}] - \delta[\varrho_A\otimes\varrho_B]\,,$$ 
that is, the amount of correlations that are lost upon considering 
the Gaussian counterpart of $\varrho_{AB}$ is equal to the amount of non-Gaussianity 
that is lost by considering the tensor product of the partial states 
$\varrho_{A}\otimes\varrho_B$ 
instead of the (correlated) bipartite state $\varrho_{AB}$. 
\par
Another quantity that is known to be maximised, at fixed covariance
matrix, by Gaussian states is the conditional
entropy \cite{GaussChannel}. Conditional entropy is defined,
given a bipartite state $\varrho_{AB}$ as 
$\S(A|B)=\S(\varrho_{AB})-\S(\varrho_B)$ \cite{Cer97}.
The proof of the extremality can be easily obtained
by means of our QRE based measure in the following way:
\begin{align}
\S(A|B) &=\S(\varrho_{AB}) - \S(\varrho_B) \nonumber \\
&= \S(\tau_{AB}) - \S(\tau_A) - (\delta[\varrho_{AB}] - 
\delta[\varrho_B]) \nonumber \\
&\leq \S_G (A|B)
\end{align}
where we used the monotonicity of the measure under partial 
trace, and where we defined the Gaussian conditional entropy 
$\S_G(A|B) = \S(\tau_{AB}) - \S(\tau_B)$ as the one evaluated by considering 
the reference Gaussian states. The two conditional entropies differ
exactly by the quantity $\Delta_1=\delta[\varrho_{AB}]-\delta[\varrho_B]$;
 thus, the more the non-Gaussianity of the overall state
is \emph{robust} under discarding a subsystem, the more
the difference between the two quantities $\S_G(A|B)$ and $\S(A|B)$ 
is near to zero. This extremality condition is important for 
several reasons:  the negative of the conditional entropy
is a lower bound for the \emph{distillable entanglement}, and 
thus, one can evaluate, given a generic bipartite state $\varrho_{AB}$,
a simple lower bound on the distillable entanglement 
based only on first and second moments of the state.
Moreover, according to the operational meaning of quantum relative
entropy given in \cite{HOW} in the framework
of quantum state merging, we observe that at fixed covariance
matrix, it is always more convenient to use non-Gaussian state.
Indeed, for positive values, the quantum conditional entropy 
quantifies how much quantum information Alice needs to send to 
Bob so that he gains the full knowledge of the bipartite state
$\varrho_{AB}$ given his previous knowledge about
the partial state $\varrho_{B}$, while for negative values 
it turns out that Alice needs to send only classical information
and moreover the two users gain entanglement to 
perform, as example, teleportation. As a consequence 
at fixed covariance matrix, by using a non-Gaussian state
you have to send less information or, for negative values,
you gain more entanglement; the gain is exactly given
by the \emph{degradation} of the non-Gaussianity
under partial trace, that is how much non-Gaussianity
is lost from the initial state $\varrho_{AB}$ tracing out
the first Hilbert space.
\par
A different interpretation of the conditional entropy has been
also given in \cite{Buscemi}, in the context of
the so-called \emph{private quantum decoupling}. It has 
been shown that,
in the limit of infinitely many copies of the initial state $\varrho_{AB}$,
the ineliminable correlations between the two parties are
quantified by the negative of the conditional entropy. Again
we have that, at fixed covariance matrix, non-Gaussian 
state have more ineliminable quantum correlations than the 
corresponding Gaussian one, and that the more nG is 
lost under partial trace operation, the more 
these quantum correlations are present.
\par
The above results lead to speculate about non-Gaussian correlations
which, being encoded in a larger set of degrees of freedom, are, at fixed
covariance matrix, higher than the Gaussian ones. In particular, we have 
observed this non-trivial connection between the robustness of nG
under partial trace and {\em decoupling} operations, as well as 
in the quantum correlations present in non-Gaussian bipartite states. 
Recently, arguments have been provided supporting the conjecture 
that at fixed energy Gaussian entanglement is the most robust
against noise in a Markovian Gaussian channel \cite{GENT}.  On the other hand, 
the same analysis have also shown that robustness of non-Gaussian states 
is comparable with that of Gaussian states for sufficiently high energy of the states.
This implies that in these regimes non-Gaussian resources can be
exploited to improve quantum communication protocols approximately over
the same distances.
\section{NG and parameter estimation}\label{s:qe}
In an estimation problem one tries to infer the value the of a parameter
$\lambda$ by measuring a different quantity $X$, which is somehow
related to $\lambda$. This often happens in quantum mechanics and quantum
information where many quantities of intereset, e.g. entanglement
\cite{EE,EEE}, does not correspond to a proper observable and should be
estimated from the measurement of one or more observable
quantities \cite{LQE}. Given a set $\{\varrho_\lambda\}$ of quantum 
states parametrized by the value of the quantity of interest, an estimator 
$\hat\lambda$ for $\lambda$ is a real function of the outcomes of the 
measurements performed on $\varrho_\lambda$. The quantum Cramer-Rao 
theorem \cite{BC94,BC96} establishes a lower bound for the variance
${\mathrm{Var}}(\lambda)$ of any unbiased estimator, i.e. for the 
estimation precision,  
\begin{align}
{\mathrm{Var}}(\lambda) \geq \frac{1}{M H(\lambda)} \label{eq:CramerRao}
\end{align}
in terms of the number of measurements $M$ and the so-called quantum Fisher 
information, which captures the statistical distinguishability of the
states within the set and itself is proportional to the Bures distance between 
states corresponding to infinitesimally close values of the parameter,
i.e. 
\begin{align}
H(\lambda) & = 4\, d^2_B (\varrho_{\lambda+d\lambda},\varrho_\lambda)
\notag \\ &= 2 \sum_{nm} \frac{\left|\langle \psi_m| \partial_\lambda
\varrho_\lambda | \psi_n\rangle\right|^2}{\varrho_n+ \varrho_m}\:,
\label{HH}
\end{align}
where we have used the eigenbasis $\varrho_\lambda = \sum_n \varrho_n 
|\psi_n\rangle\langle\psi_n |$. \par
In an estimation problem where the variation of a parameter affects the 
Gaussian character of the involved states one may expect the amount
of non-Gaussianity to play a role in determining the estimation
precision. This indeed the case: the non-Gaussianity $\delta_B$
provides an upper bound to the quantum Fisher information at fixed
covariance matrix. This is more precisely expressed by the following
\begin{tth}
If $\tau_\lambda$ is a Gaussian state and an infinitesimal variation of
the value of $\lambda$ drives it into a state $\varrho_{\lambda+d\lambda}$ 
with the same covariance matrix, then the non-Gaussianity 
$\delta_B[\varrho_{\lambda+d\lambda}]$ provides an upper bound to the
quantum Fisher information.
\end{tth}
{\bf Proof}: If $\varrho_{\lambda+d\lambda}$ and $\tau_\lambda$ have
the same CM then the nG of $\varrho_{\lambda+d\lambda}$, 
$\delta_B[\varrho_{\lambda+d\lambda}] =
\S(\varrho_{\lambda+d\lambda}||\tau_\lambda)$ equals the so-called
Kubo-Mori-Bogolubov information $\widetilde H(\lambda)$ \cite{WHH07,Ama00}, which
itself provides an upper bound for the quantum Fisher information 
$H(\lambda) \leq \widetilde H (\lambda)$ \cite{Pet96}, thus proving the theorem.
$\square$
\\ $ $ \par
The theorem says that the more non-Gaussian is the perturbed state,
the more may be distinguishable from the original one, thus allowing 
a more precise estimation.  One may wonder that when
$\varrho_{\lambda+d\lambda}$ is itself a Gaussian state
the theorem requires $H(\lambda)=0$, i.e. no reliable estimation
is possible. Indeed, this should be the case, since Gaussian states
are uniquely determined by the first two moments and thus the 
requirement of having the same covariance matrix implies that
$\tau_{\lambda+d\lambda}$ and $\tau_{\lambda}$ are 
actually the same quantum state.
\par
For situations where the CM is changed by the perturbation we have
no general results. On the other hand, it has been already shown 
that non-Gaussian states improve quantum estimation of displacement 
and squeezing parameters \cite{EKerr} and of the loss parameter 
\cite{GG09},  compared to optimal Gaussian probes.
\section{Experimentally friendly lower bounds to QRE nG}\label{s:bounds}
A drawback of the nG measure $\delta_B$ is that
its evaluation requires the knowledge of the full density matrix
$\varrho$. For this reason, it is often hard to compute when one
has only partial informations coming from some, maybe
inefficient, measurements. In literature differente approaches
have been proposed to estimate squeezing \cite{Fiu04} and 
entanglement \cite{nGEE} of Gaussian and non-Gaussian states when only
certain measurement are available in the lab.
In the following we will derive
some lower bounds to the QRE-based nG 
measure for some class of states and by considering
the possibility to perform on the states only certain efficient or
inefficient measurements.
\subsection{Diagonal state and inefficient photodetection}
\label{s:1}
\par\noindent
Let us consider a generic single-mode state diagonal 
in the Fock basis $\varrho=\sum_n p_n |n\rangle\langle n|$.
Its non-Gaussianity can be evaluated as
\begin{align}
\delta[\varrho] &= \S(\nu_N) - \S(\varrho) 
= \S(\nu_N) - H(p_n)  \label{eq:nG}
\end{align}
where $\nu_N$ is a thermal state with the same
average photon number $N=\sum_n n \, p_n$,
and from now on  $$H(p_n)=- \sum_n p_n \log p_n$$
denotes the Shannon entropy corresponding to the
distribution $\{p_n\}$.\\
Let us consider now an inefficient photodetector described
by POVM operators
\begin{align}
 \Pi_m = \sum_{s=m}^{\infty} \alpha_{m,s}(\eta) |s\rangle\langle s| \nonumber
\end{align}
with $\alpha_{m,s}(\eta)$ defined in Eq.~(\ref{eq:alphaloss}), and where $\eta$
is the efficiency of the detector. Using this kind of detection
one can reconstruct the probability distribution 
$q_m = \Tr [ \varrho \Pi_m ]$. We want to show that the 
quantity
\begin{align}
\epsilon_{\scriptstyle A}[\varrho] = \S(\nu_M) - H(q_m) \label{eq:nGP}
\end{align}
with $M=\sum_m m \, q_m=\eta N$, is a lower bound
on the actual non-Gaussianity $\delta[\varrho]$. 
To this aim we remind that an inefficient photodetection can be
described by mixing the quantum state $\varrho$
with a vacuum state at a beam splitter with 
transmissivity $\eta$ followed by a perfect photodetection 
with projective operators $P_m = |m\rangle\langle m|$. 
The corresponding probability distribution is therefore
\begin{align}
q_m &= \Tr_{12}[ U_{BS}(\eta) \varrho \otimes |0\rangle\langle 0|
 U_{BS}^{\dagger}(\eta) |m\rangle\langle m| \otimes \mathbbm{1} ] \nonumber \\
&= \Tr_1 [ \mathcal{E}(\varrho) |m\rangle\langle m|] \nonumber 
\end{align}
where $\mathcal{E}(\varrho)=\Tr_2[ U_{BS}(\eta) \varrho 
\otimes |0\rangle\langle 0| U_{BS}^{\dagger}(\eta)] $ is the 
\emph{loss} channel applied on the quantum state $\varrho$. 
Since $\varrho$ is diagonal in the Fock basis we can easily show
that $\mathcal{E}(\varrho)$ is still diagonal, 
\begin{align}
\mathcal{E}(\varrho) &= \sum_n p_n \mathcal{E}(|n\rangle\langle n|) 
= \sum_n \sum_{l=0}^n p_n \alpha_{l,n} (\eta) |l\rangle\langle l|
\end{align}
where we used that $\mathcal{E}(|n\rangle\langle n|) = 
\sum_{l=0}^n \alpha_{l,n}(\eta) |l \rangle \langle l|$.
By performing an efficient photodetection we can experimentally obtain
the probability distribution
\begin{align}
q_m &= \Tr [ \mathcal{E}(\varrho) |m\rangle\langle m|] 
= \sum_{n=m}^{\infty} p_n \alpha_{m,n}(\eta) \nonumber 
\end{align} 
The quantum state $\mathcal{E}(\varrho)$ in fully described
by $q_m$ and, by observing Eq. (\ref{eq:nG}) and (\ref{eq:nGP}) 
we can easily see that 
$\epsilon_{\scriptstyle A}[\varrho]=\delta[\mathcal{E}(\varrho)]$. 
By simply using the fact the non-Gaussianity measure $\delta[\varrho]$
is non-increasing under Gaussian maps, such as $\mathcal{E}$, we
finally obtain
\begin{align}
\epsilon_{\scriptstyle A}[\varrho] = \delta[\mathcal{E}(\varrho)] \leq \delta[\varrho].
\end{align}
This inequality tells us that performing an inefficient 
photodetection on a given quantum state $\varrho$ diagonal in the Fock
basis and if we are able to reconstruct the probability 
distribution $q_m$, we can use Eq. (\ref{eq:nGP}) to 
obtain a lower bound on the actual non-Gaussianity $\delta[\varrho]$.
\subsection{State with a thermal reference Gaussian state 
and ideal photodetection} \label{s:thermal}
Let us consider a quantum state $\varrho=\sum_{n,m} p_{n,m} |n\rangle\langle m|$
such that $\Tr[\varrho a]=\Tr[\varrho a^2]=0$. The corresponding reference 
Gaussian state is a thermal state $\nu_N$ with the same average number of 
photons $N=\Tr[\varrho a^{\dag}a]$. Let us consider now the 
quantum state $\varrho|_d$ obtained considering only the photon
number distribution $p_{n,n}$ and removing all the off-diagonal elements
that is, $\varrho |_d= \sum_n p_{n,n} |n\rangle\langle n|$. 
The reference Gaussian state of $\varrho |_d$ is again the thermal
state $\nu_N$.
Let us consider now the state
\begin{align}
\mathcal{N}_\Delta(\varrho) = \sum_{n,m} e^{-\Delta^2 (n-m)^2} 
p_{n,m} |n\rangle\langle m|. \label{eq:PhDiff}
\end{align} 
which physically corresponds to a phase-diffusion applied to 
the initial state. As a matter of fact, the same kind of evolved 
state may be obtained by the application of a random zero-mean Gaussian 
distributed phase-shift to $\varrho$, that is:
\begin{align}
\mathcal{N}_\Delta(\varrho) = \int_R d\phi \, 
\frac{e^{-\phi^2 / (4\Delta^2)}}{\sqrt{4\pi\Delta^2}} \, 
U_\phi \varrho\: U_\phi^\dag\:,
\end{align}
where $U_\phi=\exp\{-i a^\dag a \phi\}$. Upon using the invariance 
under unitary operators and the  concavity of the von Neumann entropy
one may show that
\begin{align}
\S(\mathcal{N}_\Delta(\varrho)) &\geq \int_R d\phi \, 
\frac{e^{-\phi^2 / (4\Delta^2)}}{\sqrt{4\pi\Delta^2}} \, 
\S(U_\phi \varrho U_\phi^\dag)  
= \S(\varrho)\:. \label{eq:VNEPhDiff}
\end{align}
>From 
Eq. (\ref{eq:PhDiff}) we have that 
$\varrho |_d = \lim_{\Delta\rightarrow\infty} \mathcal{N}_\Delta(\varrho)
$
and because of Eq. (\ref{eq:VNEPhDiff}) 
\begin{align}
H(p_{n,n})=\S(\varrho |_d) \geq \S(\varrho) \label{eq:VNEdiag}
\end{align}
It is straightforward to see that the non-Gaussianity
of $\varrho$ is lower bounded by the non-Gaussianity
evaluated by considering only the photon-number
distribution of the state, \emph{i.e.}
\begin{align}
\epsilon_{\scriptstyle B}[\varrho] &= \S(\nu_N) - H( p_{n,n})
\leq  \S(\nu_N) - \S(\varrho) = \delta[\varrho]   \label{eq:BoundThermal}
\end{align}
\subsection{State with a thermal reference Gaussian state and inefficient photodetection}
Let us take a quantum state $\varrho$ as in the previous 
Section, but considering the case of an inefficient photodetection .
>From the measurement, we obtain  the distribution $q_m = \Tr [\varrho \Pi_m] $,
and we can thus define a quantum state 
\begin{align}
\theta_M = \sum_m q_m |m\rangle\langle m| .
\end{align}
having $M=\sum_m m \: q_m$ average photons.
As in Sec. \ref{s:1} we have $\theta_M = \mathcal{E}(\varrho |_d)$, 
with $\varrho |_d$ defined as before and $\mathcal{E}$ denoting the loss channel. 
Then, by using the monotonicity of non-Gaussianity
under Gaussian maps, and the previous results, we obtain 
\begin{align}
\epsilon_{\scriptstyle C} [\varrho] = \S(\nu_M) - H(q_m) = 
\delta[\theta_M] \leq \delta[\varrho |_d] \leq \delta[\varrho] 
\end{align}
\subsection{Generic state with known covariance matrix and ideal photodetection}
Let us consider a generic single-mode state $\varrho=\sum_{n,m} p_{n,m}
|n\rangle\langle m|$ with a reference Gaussian state
$\tau$. Its non-Gaussianity is evaluated as 
\begin{align}
\delta[\varrho] = \S(\tau) - \S(\varrho)
\end{align}
Let us consider the quantum state $\varrho |_d=\sum_n p_{n,n} |n\rangle\langle n|$
as in the previous case and suppose that we able to evaluate its covariance matrix
of $\varrho$ (and thus the entropy of the reference Gaussian state $\tau$).
We proved before that $H(p_{n,n})=\S(\varrho| _d)\geq \S(\varrho)$ 
and thus we have the computable
lower bound on the non-Gaussianity
\begin{align}
\epsilon_{\scriptstyle D} [\varrho] = \S(\tau) - H(p_{n,n}) \leq \delta[\varrho] 
\end{align}
\subsection{Generic state with inefficient photodetection}
Let us consider a generic single-mode state $\varrho=\sum_{n,m} p_{n,m}
|n\rangle\langle m|$ with a reference Gaussian state
$\tau$. Because of the monotonicity of the measure under Gaussian maps,
we have 
\begin{align}
\delta[\varrho] \geq \delta[\mathcal{E}(\varrho)] = \S(\tau_{\eta}) - \S(\mathcal{E}(\varrho))
\end{align}
where $\tau_\eta = \mathcal{E}(\tau)$. Again, by using the inequality
derived in Eq. (\ref{eq:VNEdiag}) we obtain
\begin{align}
\epsilon_{\scriptstyle E} [\varrho] = \S(\tau_\eta) - H(q_{m}) \leq \delta[\varrho]
\end{align}
where $q_m = \Tr[ \varrho \Pi_m]$. This general lower bound can be useful
when the covariance matrix of the state can be easily derived from the 
photon number statistics of  $\varrho$ (\emph{e.g.} for phase-averaged
coherent  states).
\section{Conclusions} \label{s:conclusions}
Non-Gaussianity is a resource for quantum information processing and
thus we urge a measure able to quantify the non-Gaussian character of a
quantum state. In this paper we have addressed non-Gaussianity of states
and operations in continuous-variable systems and we have illustrated in
details two measures of nG proposed in \cite{nGHS,nGRE}, along
with their properties and the relationships between them. We used them
to assess some Gaussification and de-Gaussification processes, and in
particular we studied the role of the amount of nG in two entanglement
distillation protocols proposed in literature. The role of
non-Gaussianity appears to depend on the protocol itself, and at least
in one of the two protocols, the amount of gained entanglement at each
step of the protocol is monotonous with the nG of the initial
low-entangled state.  We have also reconsidered the extremality of
Gaussian states in terms of our measure based on the QRE, for some
relevant quantities in quantum communication, as conditional entropy,
mutual information and Holevo bound.  In particular, we found that in
the bipartite setting there is a, probably not entirely understood,
connection between correlations and nG: at fixed covariance matrix
non-Gaussian states have more correlations and this excess of
correlations is related on the amount of non-G that the quantum state
loses under partial trace operation or under decoupling.  These results,
together with recent ones on the robustness of non-Gaussian entanglement
in noisy Markovian channels implies that there are regimes where
non-Gaussian resources can be exploited to improve quantum communication
protocols. We have also seen that QRE non-Gaussianity is a bound for the
quantum Fisher information at fixed covariance matrix and thus the nG
features of quantum states may be also used to improve parameter
estimation with continouos variables.  Finally, since the evaluation of
the QRE nG measure requires the knowledge of the full density matrix,
we derive some experimentally friendly lower bounds to nG for some
class of states and by considering the possibility to perform on the
states only certain efficient or inefficient measurements.
\par
Our analysis of the properties and the applications of the two measures
of non-Gaussianity has shown that they provide a suitable quantification
of nG for the purposes of quantum information.  In particular, the QRE
based nG $\delta_B$ have several operational characterizations and it
is not too difficult to be evaluated. Since all states $\varrho$ with
the same first two moments at fixed purity have the same amount of nG
$\delta_B[\varrho]$ an interesting and remarkable picture emerges: for
many purposes the effects of nG may be described by a single global
parameter rather than ascribed to a specific higher moment.  This is
perhaps our main conclusion. Overall, our results suggest that in terms
of resources for quantum information, the amount of non-Gaussianity of a
quantum state can be evaluated using $\delta_B$, with $\delta_A$ serving
as a fine-tuning tool for specific purposes.
\par
There are several open problems requiring further
investigations about non-Gaussianity of quantum states. Among them
we mention the following ones, which also provides a summary of the unanswered
quastions posed in our paper: i) Is there any general relation between the
two measures of non-Gaussianity, at least for specific class of states? 
ii) Is there a maximum amount of distillable 
entanglement at fixed non-Gaussianity ?  iii) Which is the role of 
non-Gaussianity in parameter estimation involving a change in the 
covariance matrix?
\par
In conclusion, non-Gaussianity is a resource that can be quantified.
Our results pave the way for further development and suggest that a 
deeper understanding of the geometrical and analytical structures 
underlying the non-Gaussian features of states and operations could 
be a powerful tool for the effective implementation of quantum 
information processing with continuous variables.
\section*{Acknowledgements}
The authors would like to thank Konrad Banaszek, Carmen 
Invernizzi, Stefano Olivares, Paolo Giorda, Michele Allegra, 
Ruggero Vasile, Sabrina Maniscalco, Marco Barbieri, Alberto Porzio,
Salvatore Solimeno, Janika Paavola, Gerardo Adesso, Mauro Paternostro, 
Alessia Allevi, Maria Bondani, Francesca Beduini, Alessio Serafini, 
Jarda Rehacek, and Zdenel Hradil for useful discussions.


\begin{thebibliography}{99}
\bibitem{noteGaussianMap} 
As discussed before in Lemma B6, 
the nG measure $\delta_B$ is monotone only under full CP Gaussian maps. 
In this case we are considering
the Krauss operator corresponding to the vacuum outcomes of the measurement,
and, even if only Gaussian operations and Gaussian measurements are involved,
the inequality expressed in Lemma B6 can be violated.
\end{thebibliography}
\end{document}